\documentclass{aa}  
\usepackage{graphicx}

\newcommand{\vac}{\texttt{MPI-AMRVAC}\xspace}

\usepackage{hyperref}
\usepackage{txfonts}
\begin{document} 

   \title{First 3D radiation-hydrodynamic simulations of Wolf-Rayet winds}

   \author{N. Moens
          \inst{1}
          \and
          L. G. Poniatowski
          \inst{1}
          \and
          L. Hennicker
          \inst{1}
          \and
          J. O. Sundqvist
          \inst{1}
          \and
          I. El Mellah
          \inst{2,4}
          \and
          N. D. Kee
          \inst{1,3}
          }

     \institute{Instituut voor Sterrenkunde, KU Leuven,
              Celestijnenlaan 200D, 3001 Leuven, Belgium,\\
              \email{nicolas.moens@kuleuven.be}
         \and
            Institut de Planétologie et d’Astrophysique de Grenoble (IPAG), Université Grenoble Alpes, 38058 Grenoble Cedex 9, France 
        \and
            National Solar Observatory,
            22 Ohi'a Ku St,
            Makawao,
            Hawaii (HI),
            United States
        \and
            Center for Interdisciplinary Research in Astrophysics and Space Exploration (CIRAS), 
            Universidad de Santiago de Chile, 
            Estación Central, 
            Chile
            }

   \date{Received January xx, xxxx; accepted February xx, xxxx}

 
  \abstract
   {Classical Wolf-Rayet (WR) stars are direct supernova progenitors undergoing vigorous mass loss. Understanding the dense and fast outflows of such WR stars is thus crucial for understanding advanced stages of stellar evolution and the dynamical feedback of massive stars on their environments, and for characterizing the distribution of black hole masses.}
   {In this paper, we develop the first time-dependent, multidimensional, radiation-hydrodynamical models of the extended optically thick atmospheres and wind outflows of hydrogen-free classical WR stars.} 
   {A flux-limiting radiation hydrodynamics approach is used on a finite volume mesh to model WR outflows. The opacities are described using a combination of tabulated Rosseland mean opacities and the enhanced line opacities expected within a supersonic flow.}
   {For high-luminosity models, a radiation-driven, dense, supersonic wind is launched from deep subsurface regions associated with peaks in the Rosseland mean opacity. For a model with lower luminosity, on the other hand, the Rosseland mean opacity is not sufficient to sustain a net-radial outflow in the subsurface regions. Instead, what develops in this case, is a "standard" line-driven wind launched from the optically thin regions above an extended, moderately inflated, and highly turbulent atmosphere. We thus find here a natural transition from optically thick outflows of classical WR stars to optically thin winds of hot, compact subdwarfs; in our simulations, this transition occurs approximately at a luminosity that is $\sim 40 \%$ of the Eddington luminosity. Because of the changing character of the wind-launching mechanism, this transition is also accompanied by a large drop (on the low-luminosity end) in the average mass-loss rate. Since the subsurface opacity peaks are further associated with convective instabilities, the flows are highly structured and turbulent, consisting of coexisting regions of outflowing, stagnated, and even pockets of infalling gas. Typical velocity dispersions in our 3D models are high, 100-300 km/s, but the clumping factors are rather modest, $f_{\rm cl} \equiv \langle \rho^2 \rangle/\langle \rho \rangle^2 \sim 2$. We further find that, while the low-density gas in our simulations is strongly radiation-driven, the overdense structures are, after their initial launch, primarily advected outward by ram-pressure gradients. This inefficient radiative acceleration of dense "clumps" reflects the inverse dependence of line driving on mass density and leads to a general picture wherein high-density gas parcels move significantly slower than the mean and low-density wind material.}
   {}


   \maketitle
%

\section{Introduction}
Stars with initial masses higher than $8\, M_\odot$ play an essential role in both the chemistry and gas dynamics in galaxies such as our own Milky Way \citep{Doran2013,Ramachandran2018,Heger2003,Hopkins2014}.
At the end of their lives, many massive stars undergo supernova explosions, leaving behind compact objects such as neutron stars and black holes. Before this final event, however, 
many massive stars become completely depleted of hydrogen \citep{Crowther2007}. This depletion may happen because of strong stellar winds \citep{Conti1975, Abbott1987},  eruptive events \citep{Smith2014, Smith2017}, or binary stripping \citep{Paczynski1967,Shenar2020}. In any case, what is left will be a stripped, hot and compact Helium star. 

Such stripped stars can either be classified as classical Wolf-Rayet (WR) stars (after \citealt{WolfRayet1867}), when they are characterized by strong, broad emission lines, or they can become hot subdwarfs when they do not feature WR spectral characteristics. 
Subdwarfs are considered to be surrounded by a very thin atmosphere \citep{Han2010}. On the other hand, the strong emission features that characterize WR stars are believed to originate in a dense, radiation-driven wind that surrounds the object \citep{Beals1940}. Classical WR winds have high terminal speeds and high mass-loss rates (on the order of $10^{-7} - 10^{-4}$ $M_\odot$ $\rm yr^{-1}$ e.g., \citealt{NugisLamers2000, Hamann2019}), and are optically thick to the extent that they hide the hydrostatic stellar core. 
Linking evolutionary models of the WR hydrostatic core to spectral observations thus requires a thorough understanding of the wind dynamics \citep{Hillier1991}.
Identifying the exact launching mechanism that can produce WR winds strong enough to match observations is, however, still an active area of research (e.g., \citealt{Poniatowski2021, Sander2019}). Specifically, while the better understood winds of OB-stars on the main sequence are well reproduced by means of a line-driven wind theory assuming an optically thin continuum (originally developed by \citealt{CAK1975}, CAK), this standard theory cannot explain the high mass-loss rates inferred for high-luminosity classical WR stars \citep{Hillier1991b, Cassinelli1991, Lamers1993, Poniatowski2021}. 

Since WR stars have high luminosity-to-mass ratios, they typically exceed their effective Eddington limit, where the stellar gravitational pull is overcome by radiative acceleration already in deep subsurface layers \citep{Heger1996, Grassitelli2018}; the super-Eddington nature of these deep layers is associated with peaks in the Rosseland mean opacity, stemming from atomic recombination of iron in particular. 
\citet{NugisLamers2002} first explored this as a potential driving mechanism for WR winds. Moreover, \citet{Grassitelli2018} computed hydrodynamic, 1D, stationary models extending up to the wind sonic point. These works suggest that Galactic WR stars are driven to the sonic point by this "iron opacity bump," as well as a transition in mass loss as a function of the luminosity-to-mass ratio.
However, WR winds launched in optically thick layers through this mechanism do not meet the energy requirements to bring the stellar gas from the sonic point up to escape speed \citep{Ro2016,Poniatowski2021}.
In 1D models based purely on Rosseland opacities, this then leads to solutions that are not able to escape the stellar gravitational potential. Here, gas gets accelerated and lifted up from the hydrostatic core, before it decelerates and finally starts falling back down onto the core surface due to gravity. 
 
In \citet{Poniatowski2021} we thus suggested a hybrid opacity formulation, wherein Rosseland mean opacities are combined with the strong enhancement in line opacity that, due to the Doppler effect, is expected to occur in supersonic media (CAK). In these hybrid opacity models, gas can then get lifted up from the hydrostatic stellar core via static Rosseland mean opacity. Then, once the gas is further away from the stellar core it becomes optically thin, and efficient line driving can take over. The first 1D spherically symmetric models have been computed using this mechanism, reproducing the correct order of magnitude of mass-loss rates and providing a basic explanation for the so-called core radius problem of classical WR stars \citep{Poniatowski2021, Moens2021}. However, these 1D models still needed further ad hoc opacity enhancements to the line driving in order to prevent the mass launched in subsurface layers from falling back upon the star. 
In \citet{Poniatowski2021}, these enhancements were attributed to unknown details about the line statistics. Similarly, to avoid a nonmonotonic velocity field in 1D stationary hydrodynamic models of WR winds based on a  detailed non-local thermodynamic equilibrium (NLTE) comoving frame radiative transfer \citep{Grafener2005,Sander2019,Sander2020}, the radiation force is enhanced by first assuming transfer through a strongly clumped outflow, but then using this force to solve the equation of motion for a stationary, smooth wind (see discussion in \citealt{Bjorklund2022}, their Sect. 2.2, for some potential fundamental problems with this approach).

In a multidimensional setting, on the other hand, subsurface opacity peaks will also give rise to convective and radiative instabilities \citep{Cantiello2009, Jiang2015, Jiang2018, Goldberg2021}. Since for high-luminosity WR stars these same opacity peaks are also implied to be launching a supersonic outflow, it may be expected that, instead of being stationary, the deep WR atmosphere might consist of a complex pattern of coexisting regions of up- and down-flowing material. This would then cause the formation of structure and significant turbulent motions. 

To understand these effects, WR atmospheres and outflows need to be modeled using multidimensional, time-dependent simulation techniques, accounting properly for opacity peaks in deep subsurface layers, as well as line driving in the upper atmosphere.
Here, we present the first attempts to construct such multidimensional simulations of classical WR stars. For this, we use 
the multidimensional partial differential equation (PDE) toolkit \vac\footnote{\href{http://amrvac.org/}{http://amrvac.org/}}, a finite volume solver parallelized with the open-MPI framework \citep{Xia2018}. We apply our recently implemented modules for solving the time-dependent equations of radiation-hydrodynamics (RHD) in two and three dimensions \citep{Moens2021}. This work is thus a direct follow-up to the 1D RHD models of WR winds presented in \citet{Moens2021}. However,  rather than assuming the ad hoc boosts of the line force that were inherent to previous 1D models (see above), we now use the "Munich atomic database" \citep{Pauldrach1998, Pauldrach2001, Puls2000} to also compute a  self-consistent line opacity \citep{Poniatowski2022}.   

Section \ref{sec: Modeling} describes the applied prescription and setup for the RHD simulations, as well as the hybrid opacity model used to describe the interaction between gas and radiation. In Sect. \ref{sec: general properties}, we present the general properties of our 2D and 3D models. In Sect. \ref{sec: Grid}, we present a first exploration of the parameter space by computing models that have different input stellar luminosities, and we examine the effect this has upon the atmosphere and wind properties. Section \ref{sec: Discussion} discusses our main results and how the models could be further improved. Finally, Sect. \ref{sec: Conclusions} concludes this work with a summary and an outlook.

\section{Modeling}\label{sec: Modeling}
To model the (radiation-dominated) dynamics and energy transport in the multidimensional WR atmospheres and winds, we use the newly developed radiation-hydrodynamic module of the flexible (magneto-)hydrodynamics code \vac \citep{Xia2018,Keppens2020,Moens2021}. The formalism is fully described in \citet{Moens2021}, which also includes a number of standard benchmark cases and a test simulation of a spherically symmetric WR wind outflow. 
The RHD equations are solved on a finite volume mesh using a  "box-in--wind" approach (see \citealt{Sundqvist2018, Moens2021}). The bottom boundary of the computational domain starts deep inside the WR atmosphere at the (quasi-)hydrostatic core radius $R_{\rm c}$, and the simulations then extend several $R_{\rm c}$ into the supersonic outflowing regions, here set to $r = 6 R_{\rm c}$. In the remainder of this section, we describe some key features of this modeling framework.

\subsection{Radiation hydrodynamics}
The Euler equations of hydrodynamics describe the conservation of mass, momentum, and gas energy. With the effects of gravity and radiation included as source terms on the right-hand side, these equations are:

\begin{eqnarray}
    \partial_t \rho + \nabla \cdot (\rho \vec{v}) &=& 0 \label{eq: hd_rho}, \\
    \partial_t (\vec{v} \rho) + \nabla \cdot (\vec{v} \rho \vec{v} + p) &=& - \vec{f_g} + \vec{f_r} \label{eq: hd_mom}, \\
    \partial_t e + \nabla \cdot (e \vec{v} + p \vec{v}) &=& - \vec{f_g} \cdot \vec{v} + \vec{f_r} \cdot \vec{v} + \dot{q} \label{eq: hd_e}.
\end{eqnarray}
Here, $\rho$ is the gas density, $\vec{v}$ is the gas velocity, and $e$ is the total gas energy density, consisting of both an internal and a kinetic energy component. The source terms $\vec{f_g}$ and $\vec{f_r}$ are the forces due to gravity and radiation, and $\dot{q}$ represents the heating and cooling of the gas by radiation. The gas pressure $p$ is related to the total gas energy density $e$ via the ideal gas law, which closes the system of PDEs:
\begin{equation}
    e = \frac{p}{\gamma - 1} + \frac{1}{2} \rho v^2.
\end{equation}
In our simulations, the gas and radiation dynamics are separated where we assume the equation of state of a monoatomic, nonrelativistic and nondegenerate gas. Neglecting ionization effects, the polytropic index is appropriately set to $\gamma = 5/3$ (see also e.g., \citealt{Jiang2015}). 

The external forces due to gravity and radiation impact both the momentum and kinetic energy of the gas. In our model, the gravitational force $\vec{f_g}$ is assumed to come from a point source: 

\begin{equation}
    \vec{f_g} = \rho \frac{G M_\ast}{r^2} \vec{\hat{r}}, \label{eq: grav}
\end{equation}
which means that the WR core mass $M_\ast$ is assumed to be much greater than the mass inside the atmosphere and wind. For the models presented below, a brief order of magnitude estimate shows that the mass contained in the simulated envelope is, at maximum, six orders of magnitude lower than the mass in the stellar core. Thus, the assumption of point gravity applies in this situation.
The acceleration due to the radiation force $\vec{f_r},$ and the gas heating and cooling $\dot{q}$, depend on the radiation field. Thus, we need a formalism to treat radiation and its coupling with the gas. In this work we make use of the frequency integrated $0^{\rm th}$ angular moment equation of the time-dependent radiation transport equation in the comoving frame. This is an additional PDE that can be written in a conservative form similar to the hydrodynamic equations:

\begin{equation}
    \partial_t E + \nabla \cdot (E\vec{v} + \vec{F}) = -\dot{q} -\nabla \vec{v} : \vec{P}.
    \label{eq: rhd_E}
\end{equation}
Here $E$ is the frequency-integrated radiation energy density, $\vec{F}$ is the frequency-integrated radiation flux, and $\vec{P}$ is the frequency-integrated radiation pressure tensor. This radiation subsystem further needs a closure relation connecting the radiation flux vector, the radiation pressure tensor, and the radiation energy density scalar. 
The nonequilibrium flux-limited diffusion (FLD) approach, as described by \citet{Moens2021}, is used for this, wherein the radiation flux becomes:

\begin{equation}
    \vec{F} = \frac{-c \lambda}{\kappa \rho} \nabla E. \label{eq: fld_flux}
\end{equation}
Here the flux limiter $\lambda$ prevents the magnitude of the radiation flux from exceeding the physical limit $c E$, for the speed of light $c$, when radiation is freely streaming.  In the radiation diffusion regime, radiation follows the Eddington approximation. In the formulation above, $\kappa$ is the flux-weighted mean opacity, in units of $ \rm cm^2 \, \rm g^{-1}$. In this paper, we assume a flux-limiter in the form suggested by \citet{Levermore1981} (see also \citealt{Moens2021}). The radiation pressure tensor is then written as $\vec{P} = \vec{f} E$ for a corresponding analytic form of the Eddington tensor, following \citet{Turner2001}. The radiation force density, and the heating and cooling, are computed from the local gas and radiation quantities:
\begin{eqnarray}
    \vec{f_r} &=& \rho \frac{\kappa \vec{F}}{c} \\
    \dot{q} &=& c \kappa \rho E - 4 \pi \kappa \rho B, 
\end{eqnarray}
where $B$ is the frequency-integrated Planck function, and where we have further assumed that the energy and Planck mean opacities present in the heating and cooling terms are equal to the flux mean $\kappa$. Using the definitions $E \equiv (4 \sigma/c) T_{\rm rad}^4$ and $B \equiv (\sigma/\pi) T_{\rm gas}^4$, we can alternatively write $\dot{q}$ as a function of the radiation and gas temperatures, $T_{\rm rad}$ and $T_{\rm gas}$, namely $\dot{q} = 4 \kappa \rho \sigma (T_{\rm rad}^4 - T_{\rm gas}^4)$. Because of the smaller timescales that typically control the radiative heating and cooling terms in our setup, these are updated using an implicit method as described in \citet{Moens2021}, allowing also for nonequilibrium conditions where the radiation temperature is not necessarily equal to the gas temperature. 

\subsection{Hybrid opacity model} \label{sec: LineStat}
\begin{figure}[h]
   \centering
   \includegraphics[width=0.5\textwidth]{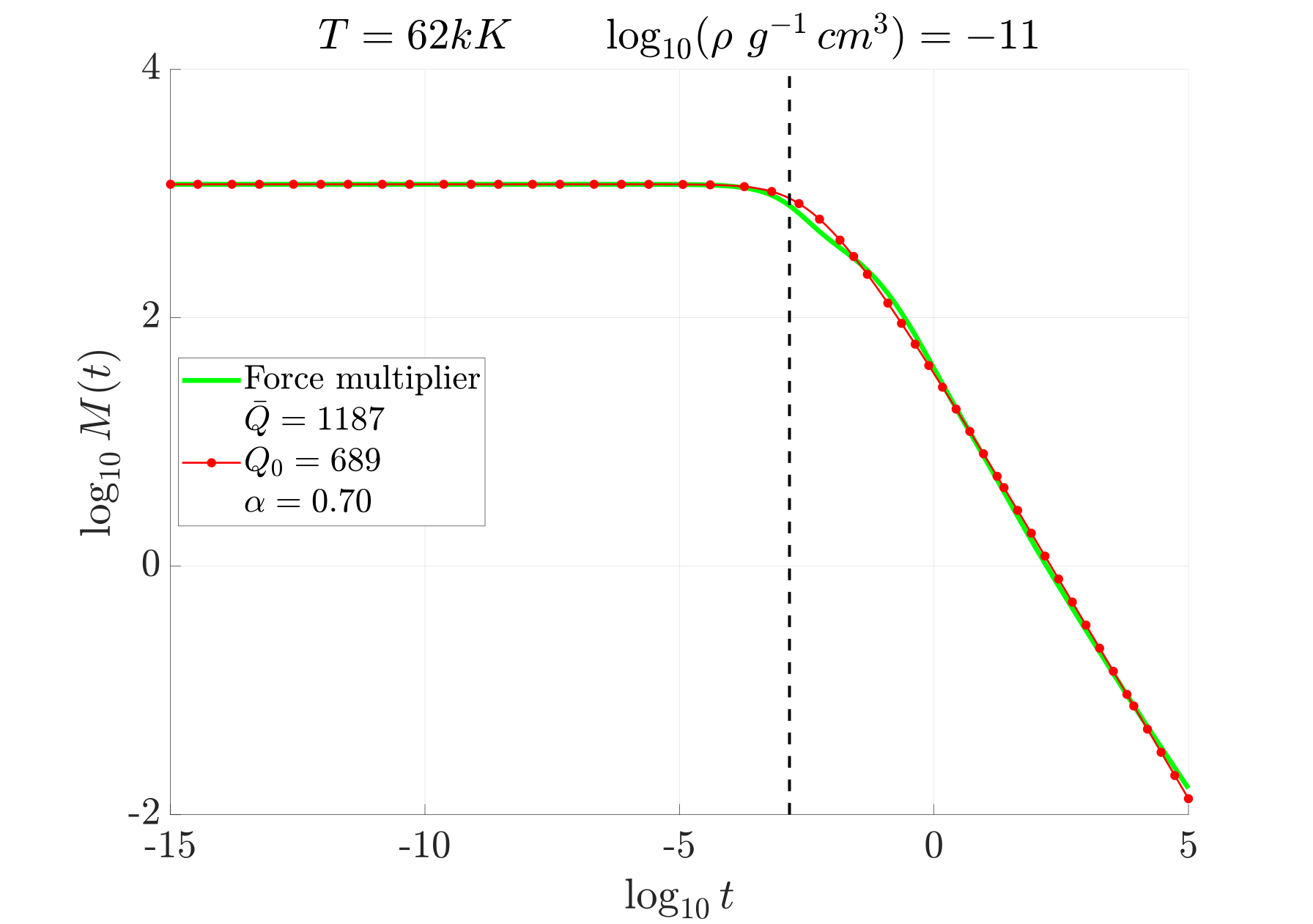}
      \caption{Force multiplier $M(t)$ as a function of the optical depth parameter $t$. The green line is obtained by calculating the line-force parameters from a line list database for a temperature of $T=62\, \rm kK$ and a gas density of $\rho=10^{-11}\, \rm g \, \rm cm^{-3}$. The red line is obtained by fitting this force multiplier with Eq. \eqref{Eq:opacity_line}. The best-fit parameters for this particular temperature and gas density are $\bar{Q}= 1187$, $Q_0 = 689$, and $\alpha = 0.70$} 
         \label{fig: Mt_fit}
   \end{figure}

%
An often-used method to obtain opacities $\kappa$ for radiation-hydrodynamics is to read them from various tabulations, for example the OPAL \citep{Iglesias1996} project, which tabulates the Rosseland mean opacity in the static limit as a function of the logarithm of the temperature $\log_{10}(T \, [K])$ and the parameter $\log_{10}(R)$ (where $R = \rho/(10^{-6}\,T\,[K])^3$), and different chemical compositions. However, previous 1D WR models (e.g., \citealt{Poniatowski2021}) have shown that these OPAL opacities, which have been gauged for static media, do not provide a good description for the total opacity when the gas becomes supersonic. 
In such layers, Doppler shifts can significantly enhance line opacities as compared to static Rosseland means, leading to efficient line driving \citep{CAK1975, Castor2004}.  
As shown in \citet{Poniatowski2021}, for classical WR stars, a strong outflow can be initiated in deep layers around $T \sim 150-200$ kK, where CAK-like line driving is quite inefficient (due to the high densities), but where a large number of bound-bound transitions in iron-like elements still contribute significantly to the Rosseland opacity (at the so-called iron opacity bump). Further out in the atmosphere, however, the temperature declines, and the static Rosseland mean opacity decreases to the extent that the initiated flow stagnates if it does not experience any additional driving (see also \citealt{Sander2019,Sander2020}). As such, building on the 1D models presented in \citet{Poniatowski2021} and \citet{Moens2021}, we here describe the total opacity as a sum of static Rosseland and CAK-like formulations
(see also suggestion by \cite{Castor2004}, their section on "velocity-stretch" opacities in Ch. 6):
\begin{align}
    \kappa = \kappa^{\rm OPAL}
    + \kappa^{\rm line},    
    \label{Eq:kap_tot}
\end{align}
where $\kappa^{\rm line}$ is the total contribution from all lines computed for a supersonic medium. In \cite{Poniatowski2022} we compute $\kappa^{\rm line}$ in the Sobolev approximation directly from a summation over the entire "Munich" line database, consisting of $\sim 4 \times 10^6$ lines \citep{Pauldrach1998, Pauldrach2001}, for a range of temperatures and densities, assuming equal radiation and gas temperatures, and local thermodynamic equilibrium (LTE). We then fit our results to a variant of the parameterization suggested by \citet{Gayley1995a}: 
\begin{equation}
    \kappa^{\rm line} = \kappa_0 \frac{\bar{Q}}{1-\alpha} \frac{\left ( (1 + Q_0 t)^{1-\alpha}-1 \right)}{Q_0 t},
    \label{Eq:opacity_line} 
\end{equation}
for 
\begin{equation}
    t = c \kappa_0 \rho \Bigl|\frac{dv}{dr}\Bigr|^{-1}, 
\end{equation}
with a fiducial normalization constant\footnote{In \citet{Poniatowski2021}, we used a normalization constant $\kappa_0 = 0.34 \, \rm g^{-1} \, cm^2$, but in these tabulations we have instead used $0.2  \, \rm g^{-1} \, cm^2$ in order to reflect the typical Thomson scattering opacity in a hydrogen-free classical WR star; this then enables a better one-to-one comparison for studies of the values of $\bar{Q}$.} $\kappa_0 = 0.2\, \rm g^{-1}cm^2$, and line-force parameters $\bar{Q}$, $Q_0$, and $\alpha$. These line-force parameters essentially represent the maximum line force in the limit that all contributing lines are optically thin ($\bar{Q}$), an effective maximum line strength ($Q_0$), and a power-law index related to the relative contributions from optically thick and thin lines ($\alpha$). From fitting Eq.  \eqref{Eq:opacity_line} to integrated opacities from our line database, we then obtain values of the line-force parameters as functions of the local temperature and density. This is quite similar to how, for example, OPAL opacity tables are constructed from computations of the Rosseland mean.  In other words, we compute and tabulate $\bar{Q}$, $Q_0$, and $\alpha$ as a function of density and temperature $\bar{Q}(\rho,T)$, $Q_0(\rho,T)$, and $\alpha(\rho,T)$. Similar tabulations have been given by \citet{Lattimer2021}, although they use a different assumed parameterization. Using these tabulations, $\kappa^{\rm line}$ is then obtained by computing $t$ from the local velocity gradient and density at each spatial point and time step in the simulation. An illustration of such a fit is provided in Fig. \ref{fig: Mt_fit}, showing the so-called line-force multiplier $M(t) = \kappa^{\rm line}/\kappa_0$ for a range of values of the optical depth variable $t$ at a fixed density and temperature. A fit (green line) of the results from the full line-list calculations (red dots) using Eq. \eqref{Eq:opacity_line} here results specifically in $\alpha =0.70$, $\bar{Q}=1187$, and $Q_0=689$, for $\rho=10^{-11} \, \rm g \, cm^{-3}$ and $T=62$ kK (typical expected density and temperature values within a WR outflow). Figure \ref{fig: Mt_fit} further illustrates how in the optically thin limit (left on the horizontal axis), $M(t) \rightarrow \bar{Q}$, and how the slope in the optically thick region (right on the horizontal axis) is indeed controlled by the $\alpha$ parameter, as discussed above (see also CAK). 

For this paper, large tables have been constructed where $\bar{Q}$, $Q_0$, and $\alpha$ are tabulated as functions of temperature and density values appropriate for the WR conditions under consideration. Specifically, we have calculated a table that covers densities in the range $\rho\in[10^{-16},\,10^{-7}]\, \rm g \, cm^{-3}$, and temperatures in the range of $T\in[10,\,100]$ kK, for a hydrogen-free plasma with the same chemical content as used for the OPAL tabulation. This means that the relative abundances of metals are the same as in the Sun, as described by \citet{Asplund2009}, and the overall metalicity $Z$ is chosen to have the solar value but without hydrogen ($X=0$), and thus $Y=1 - Z_\odot=0.98$. This can differ from other formulations such as those described by \citet{Sander2019}, where the abundance of each separate element is specified. 
These tables cover the typical mass densities expected for the WR atmosphere and wind, except for the temperatures near the lower boundary, which lay outside the table. This is because the original Munich atomic database \citep{Pauldrach1998, Pauldrach2001} only contains up to  ionization stage {\sc VIII} of the relevant line-driving elements, limiting the maximum temperature in our tables for which we have accurate atomic data. However, since the density in these high-temperature regions is also very high, this typically renders the contribution of $\kappa^{\rm line}$ ($\propto 1/\rho^\alpha$) to the total opacity small or negligible, as discussed in \citet{Poniatowski2021} and also verified here a posteriori via the lowermost panels in Fig. \ref{fig: scat_gamma}. As such, even if we were to fully neglect $\kappa^{\rm line}$ above a certain temperature threshold, the effects upon our overall results would be small. 
Nevertheless, to ensure that the total opacity does not experience any unphysically sharp transitions, we instead choose to use here  $\bar{Q}(\rho,T = 10^5 \rm K)$, $Q_0(\rho,T = 10^5 \rm K)$, and $\alpha(\rho,T = 10^5 \rm K)$ for all temperatures above $10^5$ K.\footnote{We opt for this treatment instead of an extrapolation outside the high-temperature end of the tables in order to 
ensure that the line opacity is well behaved.} 
As such, the upper limit of the tabulated temperature range should not pose any significant qualitative issues to our models. 

Using our new tabulations, we are now able to compute the spatially and time-varying line-force parameters from directly within our simulations. As such, this method constitutes a significant improvement compared to previous time-dependent radiation-hydrodynamic line-driven wind models, which typically have either assumed that these parameters are constant in both space and time, or used an ad hoc predescribed functional form \citep{Poniatowski2021}.  

\subsection{Initial and boundary conditions}
In this work, the outflows of WR stars are simulated using a box-in-wind approach. The lower boundary of the computational domain is located near the hydrostatic core such that, on average, the bottom boundary is subsonic. The outer boundary is located at a radius of $6\, R_{\rm c}$. This way, the wind can be studied from a subsonic launch all the way up to a distance at which most of the gas has reached its escape velocity.

As an initial condition, a 1D model is relaxed, as described in \citet{Moens2021}. Input parameters are the radiative luminosity $L$ at the lower boundary, stellar mass $M_\ast$, core radius $R_{\rm c}$, and a chemical composition for the OPAL and line opacity tables (see above). 
The 1D model is copied to every lateral point in the 2D or 3D box. In order to trigger initialization of structure a perturbation, sinusoidal in $r$, $y$ (and $z$), is added to the lateral momentum components. Additionally, to speed up the transition of initial conditions, the initial density profile is reduced by a factor of ten.

Boundary conditions in the lateral directions are periodic. On the bottom boundary, an extended version of our standard conditions for line-driven wind simulations is used. This means that the lower and upper boundary conditions are set as described in \citet{Moens2021}, in their Sect. 5. In summary, the mass density is kept at a fixed value,
and the momentum is extrapolated into the ghost cells. The radial component of the gradient of $E$ is set by an input bottom luminosity and the FLD closure Eq. \eqref{eq: fld_flux}. Finally, the gas energy is set to be in equilibrium with the radiation energy. Since neither the momentum nor the gas energy density is fixed at the lower boundary, this setup allows for a self-consistent computation of the mass-loss rate $\dot{M}$ and gas and radiation temperatures at the stellar core radius. 
In comparison to the 1D models in \citet{Moens2021}, for the multidimensional simulations, one complication arises in the boundary conditions for the radiation energy density. The elliptic (diffusion) term of Eq. \eqref{eq: rhd_E} is solved with a multigrid solver \cite{Teunissen2019}. For the lower boundary condition of this multigrid solver, we use a Von Neumann boundary condition, where the value of $\nabla_r E$ has to be set. In the multidimensional simulations described in this work, a single, laterally averaged value of $\nabla_r E$ is used, computed via the FLD closure Eq. \eqref{eq: fld_flux}.

Finally, at the outer boundary, $\rho$, $\bf{v}\rho$, and $e$ are linearly extrapolated outward. $E$ is set by first calculating the average optical depth of the outer boundary. For this, we assume that the portion of the wind has reached its terminal velocity, and that outside of the numerical domain, the wind continues with this constant velocity and the same mass-loss rate. Assuming only electron scattering opacity, the portion of the wind that is not simulated can now be analytically integrated to give an optical depth. This optical depth is used to calculate the radiation temperature, and from that the radiation energy density at the outer boundary is calculated.

\subsection{Finite-volume method, mesh, and spherical corrections}
\vac is a finite volume code that makes use of a quadtree or octree grid
for adaptive mesh refinement (AMR). The simulations presented in this paper were run on a Cartesian mesh, with four levels of refinement resolving the bottom boundary. On the base level, the numerical domain is only 16 cells wide and 128 cells in the radial direction, covering $0.5\, R_{\rm c}$ and $5\, R_{\rm c}$ respectively. Doubling this resolution three times to level four means that the base is refined to an effective $128$ cells laterally and $1024$ cells radially. AMR is very useful in order to resolve the, on average, subsonic layers of the wind near the core radius.  This is crucial in order to properly cover the average wind sonic point (and thus not "choke" the outflow), while still keeping the total number of cells reasonable in the outer supersonic wind. The elliptic part of the radiation energy equation is solved with a multigrid method that is integrated in \vac \citep{Teunissen2019}. However, due to the smoother operators used in the relaxation of the multigrid solution, it is currently not possible to  run these types of RHD simulations on stretched or spherical grids \citep{Briggs2000}. Thus, to correct for a spherical divergence in the RHD PDEs, we have implemented
correction terms accounting for spherical divergence effects in the fluxes of the conserved quantities, as described in Appendix A of \citet{Moens2021}.\\


\section{General properties of the multidimensional WR simulations}\label{sec: general properties}
In this section, we present the first multidimensional WR wind simulations using the hybrid opacity formalism from Sect. \ref{sec: LineStat}, which includes the improved description of line acceleration as described above. Instead of fixing $\bar{Q}$ and parameterizing $\alpha$ (as in \citealt{Poniatowski2021} and \citealt{Moens2021}), we derive local line-force parameters as computed from the line statistics of an atomic database. This means we do not choose a fixed set of line-force parameters a priori (which is somewhat arbitrary and can have a significant effect on the model dynamics), but instead we compute and update them locally as the simulations evolve. 
Our wind models depend on: stellar mass, core radius, stellar luminosity, and a chemical composition for the opacity tables. The input parameters of the generic (2D and 3D) models discussed in this section are summarized in Table \ref{table: input}. The values for stellar mass ($M_\ast = 10 \, M_\odot$) and radius of the hydrostatic core ($R_c = 1\, R_\odot$) were taken from the 1D model described in \citet{Poniatowski2021}, which were inspired by a calculation of the helium main sequence using the stellar evolution code MESA \citep{Paxton2019} (see also the $M_\ast = 10 \, M_\odot$ models in \citet{Grassitelli2018}). These values are on the low end of the mass-radius regime modeled by \citet{Langer1989}, and just below but in line with models by \citet{Maeder1987}.
In Table \ref{table: input}, the radiative stellar luminosity is expressed in units of the electron scattering Eddington luminosity $L_{\rm Edd} \equiv 4 \pi G M_\ast c/\kappa_e$, for Thomson scattering opacity $\kappa_e = 0.2 \, \rm cm^2 g^{-1}$, appropriate for our hydrogen-free simulations. In the next section (Sect. \ref{sec: Grid}), the Eddington ratio $L_\ast/L_{\rm Edd}$ is varied in four steps by changing the stellar luminosity to examine how this affects the character of the simulated outflows and the resulting structures. 

\begin{table}
\caption{Input stellar parameters and gas mass fractions for the generic 2D and 3D models discussed in Sect. \ref{sec: general properties}.}             
\label{table: input}      
\centering                          
\begin{tabular}{rl}        
\hline\hline                 
parameter & value \\    
\hline                        
  $M_\ast$ & $10 \, M_\odot$ \\
  $R_{\rm c}$ & $1 \, R_\odot$ \\
  $L_{\ast}$ & $0.67 \, L_{\rm Edd}$ \\
             & $10^{5.64} \, L_\odot$ \\
  $\kappa_{e,0}$ & $0.2 \, cm^2 \, g^{-1}$ \\
  $X$, $Y$, $Z$ & $0$, $0.98$, $0.02$ \\
\hline                                   
\end{tabular}
\end{table}

\subsection{2D model with self-consistent line force} 
\label{sec: SimulationResults}

\subsubsection{Thermal and dynamical timescales}
Structures in the simulated wind have typical velocities on the order of $10^8  \, \rm cm\, \rm s^{-1}$, which sets a characteristic dynamical timescale $\tau_{\rm dyn}$, here defined as $\tau_{\rm dyn} = R_{\rm c} / (10^8 \, \rm cm\, \rm s^{-1})$. This is the timescale at which we see changes in the positioning and shapes of over-densities and filaments over a spatial extent of about $1 R_{\rm c}$. The readjustment of the entire atmosphere (and wind) as a whole also depends on its thermal timescale $\tau_{\rm th}$, which can be estimated following \cite{Grassitelli2016}: 
\begin{equation}
    \tau_{\rm th} = \frac{G M_\ast M_{\rm env}}{R_{\rm c} L_\ast} \label{eq: tau_th}.
\end{equation}

Here, $M_{\rm env}$ is the mass contained in the atmosphere surrounding the WR core. By integrating over the average radial density profile of the simulation, we obtain the envelope mass, and using Eq. \ref{eq: tau_th}, the thermal timescale of the atmosphere is estimated to $\tau_{\rm th} \approx 300 \, {\rm s} \approx 0.5 {\tau_{\rm dyn}}$. Since the thermal and dynamical timescales are on the same order, in the rest of this paper, we will only refer to the dynamical timescale when discussing the evolution of the winds.

\subsubsection{Density and radial velocity maps} \label{sec: rho v maps}

\begin{figure*}[t]
\centering
\includegraphics[width=\linewidth]{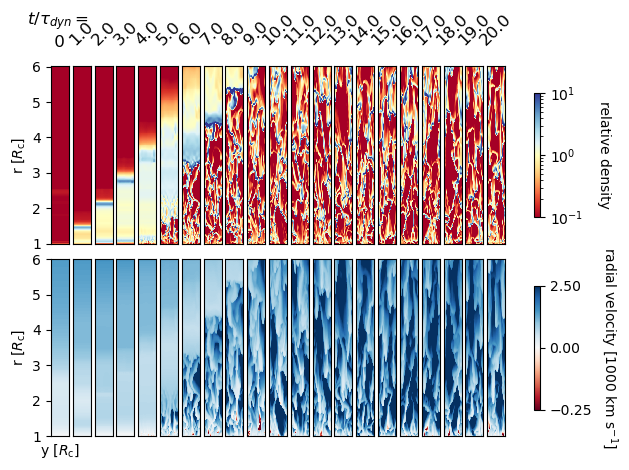}
  \caption{Color map of the relative density (top panel) and the radial velocity (bottom panel), for consecutive snapshots of the 2D model (Table \ref{table: input}) at every $\tau_{\rm dyn}$. The figure shows the breakup from initial conditions and the formation of structures in the regions close to the core, which are then carried outward.}
     \label{fig: timeseries_rho_v}
\end{figure*}

Here and throughout the paper, a relative density of the gas is used to characterize when gas in the wind is either in a clump or filament, or in the low-density medium in between overdense structures. The relative density is calculated by dividing the density at each point by an average density at the same radius. This average density is computed by averaging over the lateral coordinate over several snapshots, covering 40 dynamical timescales, well after the passing of the initial conditions. 
The upper and lower panels in Fig. \ref{fig: timeseries_rho_v} show color maps of the relative density and the radial velocity for a selected sample of snapshots of the 2D WR wind model (Table \ref{table: input}), separated by one $\tau_{\rm dyn}$ each. The upper panel of the figure allows us to focus on the structural characteristics of the gas and its evolution.    

The figure shows how the initially smooth outflow is broken up, leading to a dynamically active wind with extensive structure formation in both density and velocity. In the top panels of Fig. \ref{fig: timeseries_rho_v}, the relative density shows the formation of higher density filaments close to the lower boundary, 
which are then accelerated outward with velocities slower than the surrounding less dense material (see Sect. \ref{sec: clump dyn} below). The lower panel further shows that close to the bottom of the simulation, these higher density regions also sometimes have negative radial velocities, indicating  material that is falling back onto the hydrostatic core. 

   \begin{figure*}[t]
    \centering
    \includegraphics[width=\linewidth]{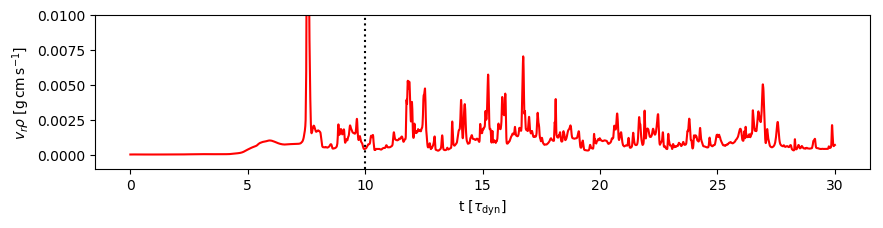}
      \caption{
      Mass flux through the $5\, R_{\rm c}$-plane of the 2D  (Table \ref{table: input}) simulation as a function of time. $5\, R_{\rm c}$ is close to the outer boundary, but still sufficiently removed to avoid effects due to boundary conditions. In the figure, the mass flux stabilizes after $\sim 10 \tau_{\rm dyn}$.} 
         \label{fig: 2D_mdot}
    \end{figure*}

Figure \ref{fig: 2D_mdot} shows, as a function of time, the mass flux\footnote{It is important to note that the time variation of this mass flux is by no means representative of the time variation of the stellar mass-loss rate in our simulations; variation of the latter will be substantially lower due to averaging effects over the full 4$\pi$ sky.} through the simulation near the outer edge of the simulation, at $5 R_{\rm c}$. Through inspection of both Fig. \ref{fig: timeseries_rho_v} and Fig. \ref{fig: 2D_mdot}, it can be seen that after $\approx 10 \tau_{\rm dyn}$, the simulation is no longer affected by the initial conditions.

\subsubsection{Origin of the structure}
We note that the structures seen in these simulations do not arise from the line-driven instability (LDI, e.g., \citealt{Owocki1988}; \citealt{Sundqvist2018}), which is in any case expected to be damped in optically thick regimes \citep{Gayley1995}. Instead, they are a result of instabilities related to the opacity peak associated primarily with iron recombination (often called the "iron opacity bump") mentioned above (see also \citealt{Jiang2015}; \citealt{Jiang2018}). 
Indeed, it can be shown that when a stellar envelope dominated by radiation pressure approaches the Eddington limit, the standard Schwarzschild criterion for convection will be fulfilled \citep[e.g.,][]{Langer1997}. 
Moreover, \citet{Castor2004} used a simple Kramer-type opacity law in a linear perturbation analysis to show that a star approaching the Eddington limit can get an imaginary Brunt-V\"ais\"al\"a frequency (see his Eq. 7.48), and thus produce an absolute instability. 
In our models, the Eddington ratio across the iron opacity bump increases to above unity, and the atmosphere becomes convectively unstable; in Appendix \ref{App: BruntVaisala} we discuss briefly how structure formation in our simulations may relate to the linear analysis by \citet{Castor2004}. 
It is important, however, to also distinguish the structures here from the well-studied convective motions occurring in subsurface layers of stars like our Sun (e.g., \citealt{Stein1998}). In such low-luminosity stars, radiation pressure plays a limited role and the observed convective motions are typically slow and subsonic. By contrast, the high-luminosity WR simulations presented here are radiation dominated and energy transport by means of convection is inefficient \citep[e.g.,][]{Grafener2012}. 
This means that radial motions that are highly supersonic with respect to the gas sound speed are initiated from the subsurface layers (in cases where the Eddington ratio is high enough), which then break up into a very turbulent flow where supersonic fast and slow streams of gas coexist at different lateral positions. Here, one may view gas parcels that are initiated but not able to reach their local escape speed (and thereby stagnate or even start falling back upon the core), as localized regions where the wind fails to escape the stellar gravitational potential, existing within the general multidimensional flow structure. 
After the initial break-up, the characteristic density structures observed in the simulations soon develop into patterns of high-density, finger-like filaments. Close to the core, these density filaments are mainly oriented in the 
radial direction.


\subsubsection{Clump dynamics} \label{sec: clump dyn}
\begin{figure}[h]
   \centering
   \includegraphics[width=0.5\textwidth]{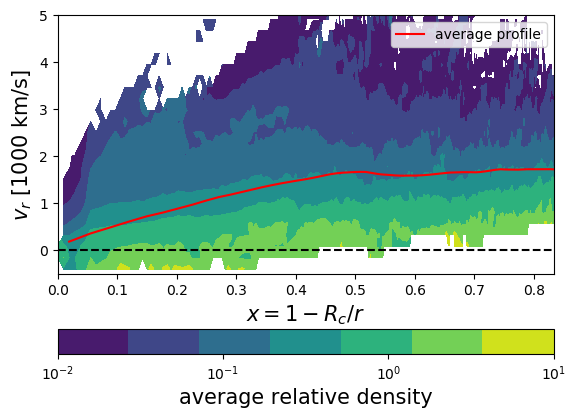}
      \caption{
      Average relative density for each radial velocity at different radii in the 2D model  (Table \ref{table: input}). Yellow indicates that a gas with a given velocity and radius is denser than the average gas at that radius, while dark blue indicates that the gas is less dense. In red are the time and laterally averaged radial velocity as a function of radius.
      }
         \label{fig: scat_v}
   \end{figure}

Figure \ref{fig: scat_v} further displays the radial velocity of the wind as a function of radius, scaled to $x = 1-R_{\rm c}/r$. In the plot, data from each of the cells for 40 snapshots are binned in discrete bins in ($r$,$v_r$)-space. Then, each bin is color-coded with the average relative density in that bin:
\begin{equation}
    \tilde{\rho}(r',v_r') = \left<\frac{\rho(r=r',v_r=v_r')}{\left<\rho(r=r')\right>}\right> .
\end{equation}
Yellow parts in the plot thus represent a gas that has a density higher than the average density at that radius. Dark blue parts represent a density lower than the average density at that radius. The 40 snapshots used in the analysis for Fig. \ref{fig: scat_v} are taken $0.5 \tau_{\rm dyn}$ apart, starting at $t=10 \tau_{\rm dyn}$.

Overplotted on the color map is the average radial velocity profile, taken from averaging over all lateral cells in those 40 snapshots. From this figure, we note that there is a large spread, of up to three orders of magnitude, in densities present at any given radius. Also, there is a clear anticorrelation between the average relative density and velocity; the high-density, clumped material is significantly slower than the average velocity, whereas the low-density material typically flows much faster. In these specific snapshots, negative velocities are also present up until $r \approx 1.5 R_{\rm c}$ ($x = 1- R_{\rm c}/r \approx 1/3$).

\begin{figure}[h]
   \centering
   \includegraphics[width=0.5\textwidth]{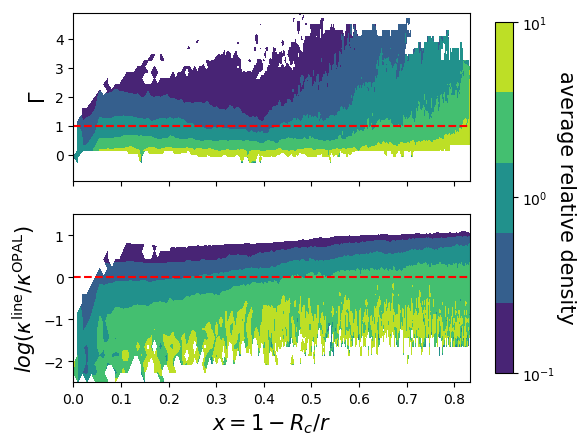}
      \caption{
            Average relative density for each Eddington factor (top panel) or ratio of opacities (bottom panel) at different radii in the 2D model (Table \ref{table: input}). Yellow indicates that a gas with a given Eddington factor or opacity ratio and radius is denser than the average gas at that radius, while dark blue indicates that the gas is less dense.}
         \label{fig: scat_gamma}
   \end{figure}

Figure \ref{fig: scat_gamma} uses the same method as Fig. \ref{fig: scat_v}, but now displaying the average relative densities in the ($r$,$\Gamma$)-plane in the top panel, with the Eddington parameter $\Gamma$:

\begin{equation} 
    \Gamma \equiv f_r/f_g, 
\end{equation} 
%
and in the bottom panel, the average relative density in the ($r$,$\kappa^{\rm line}/\kappa^{\rm OPAL} $)-plane, where $\kappa^{\rm line}/\kappa^{\rm OPAL} $ quantifies the relative importance of the two opacity sources in the hybrid opacity model (Eq. \ref{Eq:kap_tot}).
%
To be able to radiatively drive gas out of the gravitational potential, the Eddington parameter should be $\Gamma > 1$. However, as illustrated in Fig. \ref{fig: scat_gamma}, many of the high-density clumps (color-coded as yellow in the plot) 
are not super-Eddington (that is, they have $\Gamma < 1$). This low radiative acceleration experienced by the high-density material is caused by the inverse dependency of the line force on density; for $Q_0 t \gg 1$ the scaling is $\kappa^{\rm line}  \sim 1/\rho^\alpha$ (see Eqs. 11-12). This is further emphasized by the lower panel of Fig. \ref{fig: scat_gamma}, which demonstrates the inefficiency of line driving for a high-density material by displaying the ratio $\kappa^{\rm OPAL}/\kappa^{\rm line}$.  

We note that close to the stellar core, the high temperatures lie outside of the tables of the line-force parameters. As explained in Sect. \ref{sec: LineStat}, in our simulation, $\alpha$, $\bar{Q}$, and $Q_0$ are approximated by assuming that they do not change significantly with temperature outside of the tabulation domain. From the bottom panel of Fig. \ref{fig: scat_gamma}, it can be seen that $\kappa^{\rm line}$ is enhanced by $\sim 10 \%$ of $\kappa^{\rm OPAL}$ for a relatively low-density material, and $\sim 1 \%$ of $\kappa^{\rm OPAL}$ for a relatively high density material. In future work, it will be important to expand upon the tables to incorporate self-consistent line-force parameters also at higher temperatures. For this, the line database needs to be complemented with data for the higher ionization levels of the important metals, which provide the lines at such temperatures. 

\begin{figure}[h]
   \centering
   \includegraphics[width=0.5\textwidth]{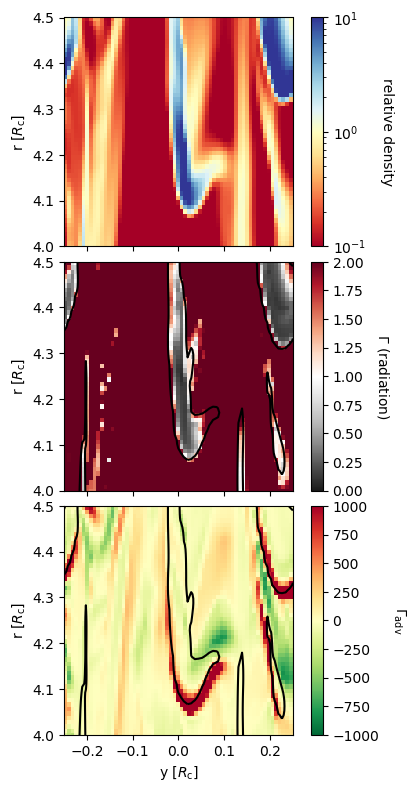}
      \caption{Momentum balance around a sub-Eddington clump in our 2D model (Table \ref{table: input}). The top panel shows the relative density where the up direction in this plot is radially outward. The bottom two panels show the Eddington factor and the ram-pressure gradient scaled to gravity, with iso-contours of relative density in black indicating the approximate location of the clump. The force distribution around this clump is characteristic of most structures observed in our simulations.}
         \label{fig: ram_p}
   \end{figure}

In the absence of efficient line driving, and from a purely radiation-gravitation force balance, one might expect these clumps to decelerate and even fall back onto the stellar surface at all radii \citep[e.g.,][]{Poniatowski2021}, and not only close to the surface as indicated by Fig. \ref{fig: scat_v}. However, closer inspection of the force balance, including the advection of momentum into the clumps, reveals that the clumped regions seem rather to be advected outward by ram pressure. In other words, the low-density regions are very efficient at picking up momentum via line acceleration (in these regions $\Gamma > 1$), accelerating this gas to high velocities before it crashes into the star-side face of high-density clumps. Momentum is then transferred between the low- and high-density regions without the need for direct line driving on the latter. This can also be seen directly from the momentum Eq. \eqref{eq: hd_mom}. Just as $\Gamma$ is defined as the ratio of radiative to gravitational force, a similar dimensionless quantity can be defined for radially advected momentum:
\begin{equation}
    \Gamma_{\rm adv} \equiv \frac{- \nabla \cdot (v_r \rho \vec{v})}{f_g} = \frac{- [\nabla \cdot \vec{P}_{\rm ram}]_r}{f_g},
\end{equation}
where the second equality introduces the tensorial ram pressure as the outer product between the momentum and velocity $\vec{P}_{\rm ram} = \vec{v} \rho \vec{v}$. In the right hand side of the equation above, $\Gamma_{\rm adv}$ is then defined as the radial component of the divergence of the ram-pressure tensor, which is then again a scalar, scaled with the radial gravitational acceleration.
If we now ignore the gas pressure term, which is small in these supersonic regions, the radial component of the momentum equation can be rewritten as:
\begin{equation}
    \partial_t (v_r \rho) = f_g (\Gamma + \Gamma_{\rm adv} - 1).
\end{equation}
%
This equation shows that the experienced acceleration does not only depend on $\Gamma$, but instead on the sum of $\Gamma$ and $\Gamma_{\rm adv}$. Figure \ref{fig: ram_p} shows this effect, comparing $\Gamma$ and $\Gamma_{\rm adv}$ around a representative gas clump of high density. Here, it is shown that the clump itself is not line driven, but momentum is rather added to the clump via ram pressure acting on its bottom edge.

\subsubsection{Line-force parameters}

\begin{figure}[h]
   \centering
   \includegraphics[width=0.5\textwidth]{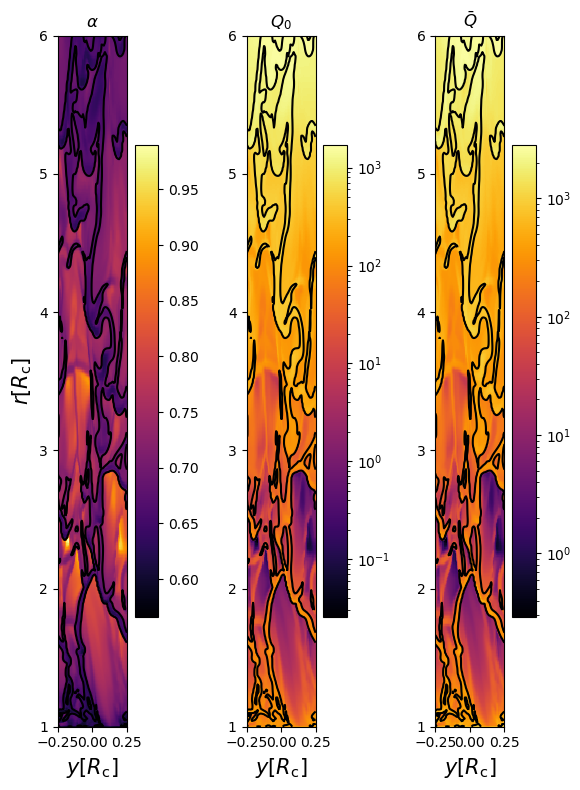}
      \caption{Color map of the line-force parameters. From left to right, the color in the panels represents the values for $\alpha$, $Q_0$, and $\bar{Q}$ for a snapshot in the 2D simulation (Table \ref{table: input}). The clumped structure is indicated by a black iso-contour marking a relative density of one.}
         \label{fig: cmap_aqq}
   \end{figure}

The acceleration experienced by the gas also depends on the locally computed and varying line-force parameters $\alpha$, $\bar{Q}$, and $Q_0$ (see Sect. 2). Figure \ref{fig: cmap_aqq} shows color maps of these parameters, again computed from the 2D model discussed above at a snapshot taken well after the relaxation of the initial conditions. The figure illustrates that the values of the line-force parameters, for the most part, reside in the typical ranges discussed by \citet{Poniatowski2022}. For example, $\alpha$ here is quite well constrained within a range $\ga 0.5$ and $< 1$. Moreover, overdense structures tend to have somewhat lower values of $\alpha$, as illustrated by the enclosed black contours in Fig. \ref{fig: cmap_aqq}. This is consistent with the general discussion in \citet{Poniatowski2022}, who find that for large parts of the ($T$, $\rho$) space, $\alpha$ indeed tends to slightly decrease with increasing density. The variations in $\bar{Q}$ and $Q_0$ are somewhat larger, ranging from the typical $\bar{Q} \sim Q_0 \sim 10^3$ to much smaller values of order ten or so. Overall, though, we find that the general inverse density dependence $\kappa^{\rm line} \sim 1/\rho^\alpha$, as discussed above, has a larger effect on the generic driving characteristics of the structured WR outflow than the temporal and spatial variation of the line-force parameters. 
Nonetheless, the large local variations in line-force parameters that we find, emphasize the importance of relying on a locally computed opacity rather than using pre-computed and fixed values for these parameters.

\subsubsection{Temperature profile}

\begin{figure}[h]
   \centering
   \includegraphics[width=0.5\textwidth]{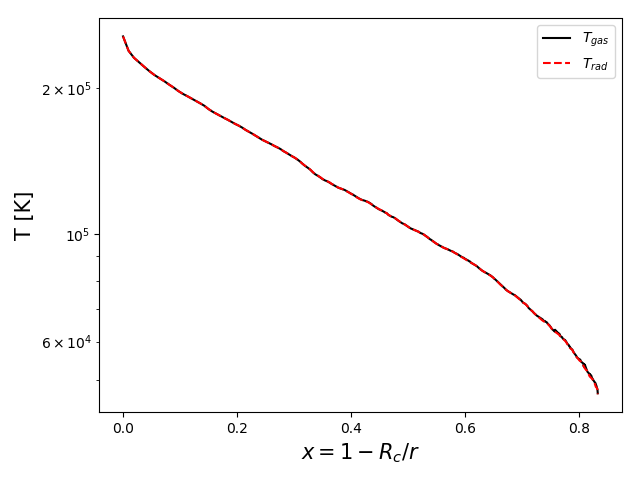}
      \caption{Comparison of the average gas and radiation temperatures in the 2D model (Table \ref{table: input}).}
         \label{fig: TradTgas}
   \end{figure}

Figure \ref{fig: TradTgas} shows the temporal and lateral averaged gas and radiation temperatures as a function of the scaled radial coordinate $x = 1 - R_{\rm c}/r$. The figure illustrates that the typical average temperatures seen in 1D models \citep{Poniatowski2021,Moens2021} are overall well preserved also within our multidimensional simulations. It further illustrates that the average gas and radiation temperatures are almost identical. In our simulations, this arises from efficient radiative cooling of shock-heated regions in the dense outflows, meaning the gas quickly settles down to almost radiative equilibrium conditions with $T_{\rm rad} \approx T_{\rm gas}$. Closer inspection of specific snapshots indeed reveals that there are quite a few gas parcels with $T_{\rm gas}$ significantly higher than $T_{\rm rad}$, which means that shocks present in the simulations are essentially isothermal. 
This may also reflect a quite general issue of resolving such shock-heated layers in supersonic line-driven flows (see discussion in \citealt{Lagae2021}). 

\subsection{Comparison between 2D and 3D models}
\label{sec: 2Dv3D}

\begin{figure}[h]
   \centering
   \includegraphics[width=0.5\textwidth]{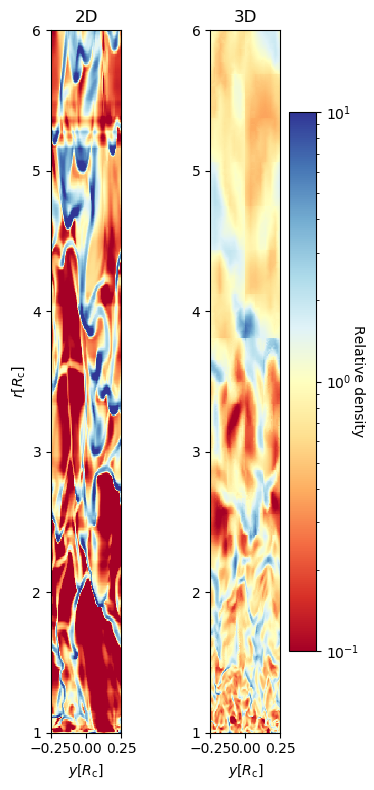}
      \caption{Relative density of the base model (Table \ref{table: input}), comparing results from a 2D setup to a 3D setup. On the left is a snapshot of the 2D model and on the right a 2D slice from the 3D model at a fixed $z$-coordinate. Both snapshots are taken well after numerical relaxation and are representative of the respective simulations.
}
         \label{fig: relrho_2D_3D}
   \end{figure}

\begin{figure*}[ht]
    \centering
    \includegraphics[width=\linewidth]{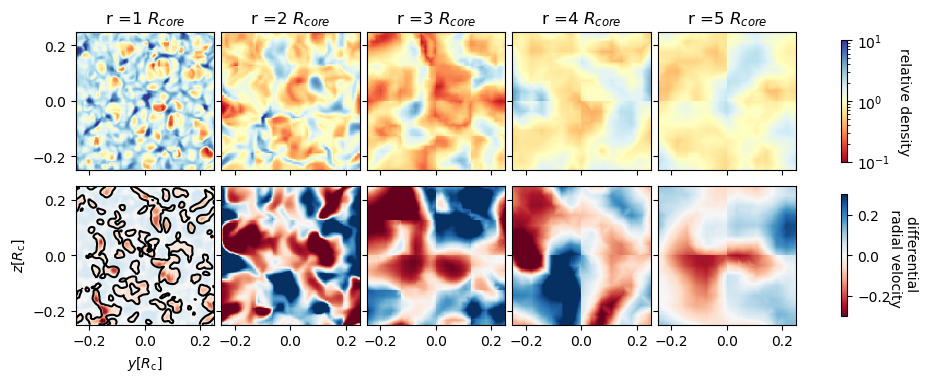}
      \caption{Relative density and radial velocity for a representative snapshot of the $\Gamma 3$ 3D simulation. In the top row, overdensities are indicated in blue and underdensities are indicated in red. In the bottom row, gas moving faster than average is indicated in blue, and gas moving slower than average is indicated in red. Different columns represent different radial coordinates, with the first column representing a radius just above the core radius. In the bottom row, the black contour represents the region where the absolute radial velocity is equal to zero.} 
         \label{fig: lat_slice}
    \end{figure*}

As discussed above in Sect. \ref{sec: rho v maps}, once developed, the density structure emerging in the simulated winds consists of high-density filaments, preferentially oriented in the radial direction. 
These characteristic structures are observed in both 2D and 3D simulations, as shown by the relative density maps  
displayed in Fig. \ref{fig: relrho_2D_3D}. Both plots shown in the figure are from a single snapshot, at a time well after numerical relaxation of the initial conditions. 
For the 3D case, a slice was taken at a fixed position in the second lateral ($z$-)direction. 
The figure shows that, although the overall characteristic structure persists, the relative density profile is sharper in 2D than in 3D, and the contrast between over- and under-dense regions is also higher. 
This can be explained by the fact that in 3D, the structures have an additional transverse dimension over which they can break up. Effectively, this means that in 3D there are more paths for the high-density gas to spread out than in 2D, leading to a smoothing effect on the overall structure. Although not shown in the figure, similar results are also obtained when inspecting the lateral velocity and temperature profiles in 2D vs. 3D. 
Such a break-up of structures in the second lateral dimension is illustrated in Fig. \ref{fig: lat_slice}.
Here, the upper panel shows lateral slices of the relative density at different radii above the core radius. 
The structures formed in the 3D box appear fairly isotropic in the lateral plane, rather than filaments with a preferred lateral direction.
Moreover, as we move away from the stellar core, these structures tend to grow in size as they move outward in radius along with the highly supersonic mean flow of the gas. The lower panel then displays corresponding slices of radial velocity. Due to the large range of velocities present, we normalize the slices to a corresponding mean velocity (computed by an average over the lateral slice). Blue portions thus correspond to velocities above the average at that radius, and red to those below. In addition, a black contour is added to mark regions where the absolute radial velocity is zero, encircling regions of negative velocity. As was also seen in Fig. \ref{fig: scat_v}, this illustrates that negative velocities are only present in layers quite close to the core (left-most plot), and that in all outer layers the simulation indeed can be characterized as an outflowing wind with a large velocity dispersion.     



\section{From hot subdwarfs to WR stars} \label{sec: Grid}
Starting from the simulations presented above, in this section, we calculate additional models by varying the bottom boundary luminosity flowing into the simulation box while keeping the stellar mass and core radius fixed. This serves to vary the basic scale of radiation-$\Gamma$, allowing us to investigate the conditions under which a radial wind outflow can already be launched from the subsurface iron opacity bump, and also the conditions under which the launched wind then can be sustained to high radii. Table \ref{table: gridresults} shows the basic parameters for the simulated stars in our grid, with models conveniently parameterized according to their Eddington luminosity $L_{\rm Edd}$.
The core radii $R_{\rm c}$ and masses $M_\ast$ are identical for all models, and are set to the same values as in Table \ref{table: input}. Here, we note that the stellar luminosity $L_\ast$ in the "dynamically inflated" 1D WR model by \citet{Poniatowski2021} lies approximately between the two models with the lowest luminosities in Table \ref{table: gridresults}, while the core radius and stellar mass are the same. It should be noted that the simulated models have luminosities slightly above the predicted $M_\ast,L_\ast$-relation given by, for example, \citet{Langer1989}. 

All simulations presented in this section are computed in full 3D (for discussion on basic differences between 2D and 3D models, see Sect. \ref{sec: 2Dv3D}). The resulting parameters obtained from the simulations are summarized in Table \ref{table: gridresults}.

As discussed in this section, varying the value of $L_\ast/L_{\rm Edd}$ allows us to discover and analyze a natural transition from high-luminosity stars with WR-type dense (optically thick) outflows, to hot, compact subdwarfs with less dense (optically thin) line-driven winds. 
Although He-stars can span a range in $\Gamma_e$, due to their tight mass-luminosity relation, it is unlikely that all models presented here (four different luminosities with one and the same mass and radius) would all be the result of realistic stellar evolution. In this work, we are mainly concerned with investigating the possible effect of different electron scattering Eddington ratios on the wind dynamics.

\begin{figure*}[t]
\centering
\includegraphics[width=\hsize]{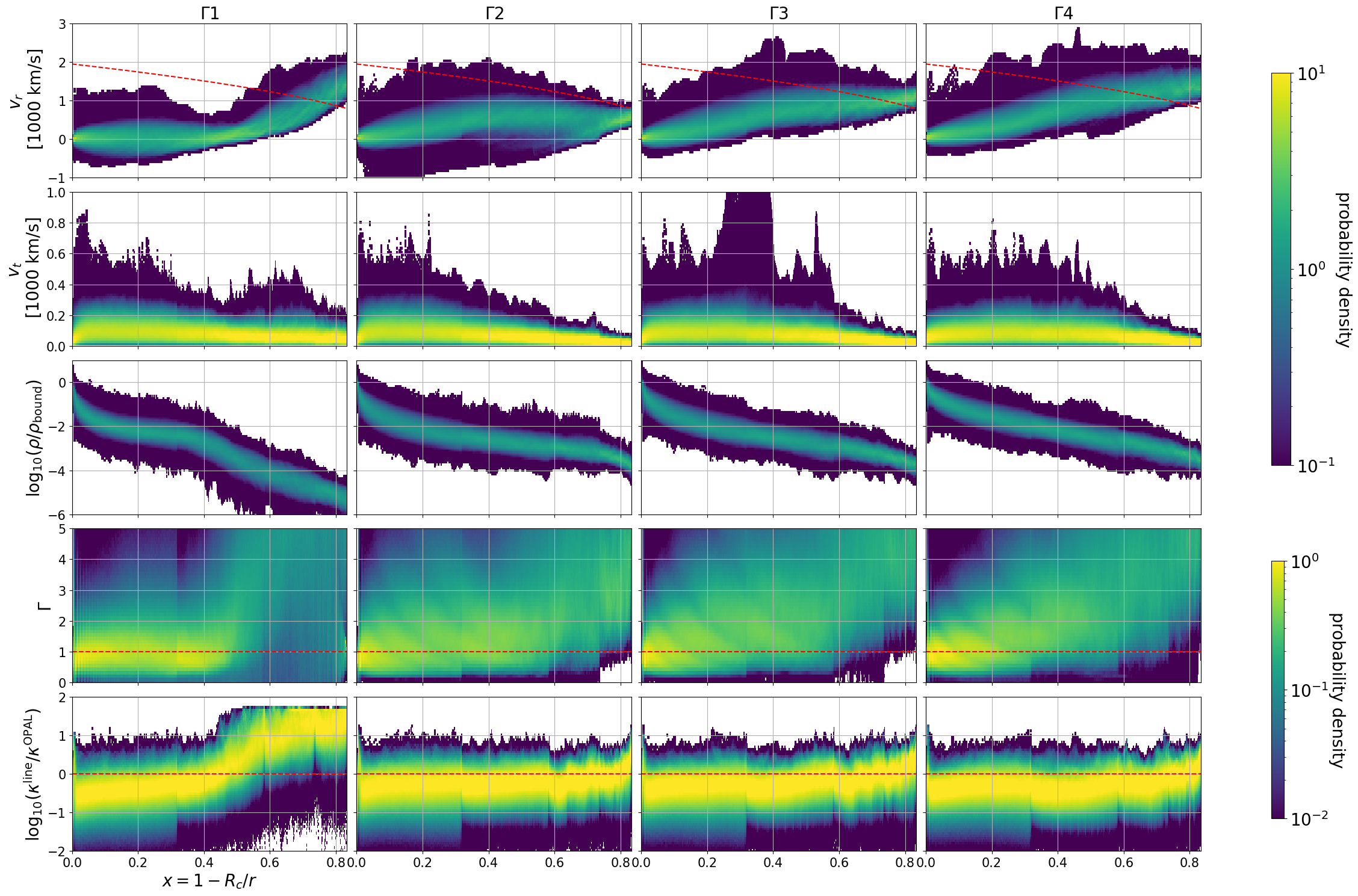}
  \caption{Probability densities for different quantities. Lateral slices from ten snapshots are used to calculate a probability distribution function at each radial point. The ten snapshots are taken after the numerical relaxation of the initial conditions and are all $1 \tau_{\rm dyn}$ apart. Different columns represent models with different boundary luminosities, increasing the luminosity from left to right. The different rows represent, from top to bottom: the density-weighted radial velocity, the magnitude of the tangential velocity $v_t = (v_y^2 + v_z^2)^{1/2}$, the gas density, the effective Eddington parameter, and the ratio of $\kappa^{\rm line}$ over $\kappa^{\rm OPAL}$. The color maps in the top three rows are scaled differently to those of the bottom two rows for visibility reasons. This is because the probability distribution functions of the top three quantities are typically thinner and higher peaked. The probability density functions are normalized such that the integral over the variable shown on the y-axis gives unity at each radius.} 
     \label{fig: prof_grid}
\end{figure*}

\begin{figure}[h]
 \centering
\includegraphics[width=0.5\textwidth]{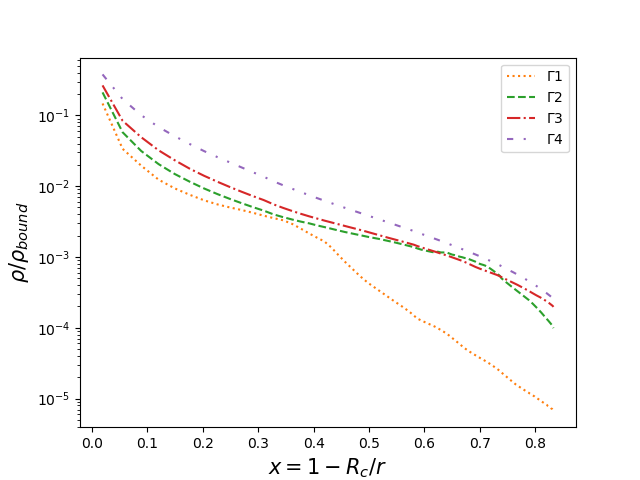}
  \caption{For each of the four 3D models, this figure shows the average density profile as a function of radius. 
}
     \label{fig: 1D_density_prof}
\end{figure}

\begin{figure}[h]
\centering
\includegraphics[width=0.5\textwidth]{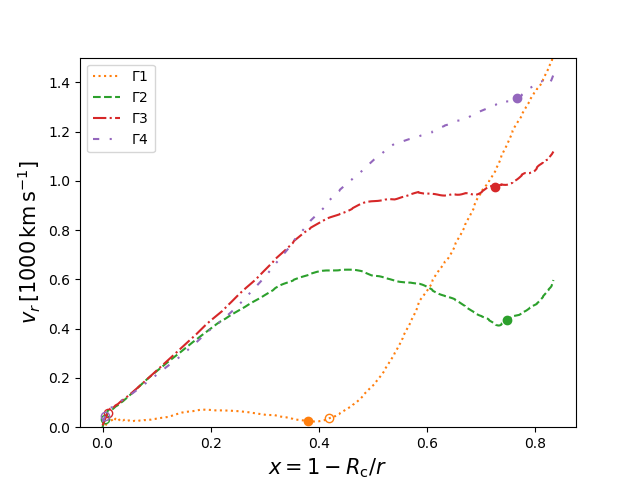}
  \caption{For each of the four 3D models, this figure shows the average velocity profile as a function of radius. The average sonic point has been indicated with an empty circle, and the mean photospheric radius with a filled circle.}
     \label{fig: 1D_velocity_prof}
\end{figure}

\begin{table*}
\caption{Results for the grid of 3D models with varying input luminosities, and a fixed $M_\ast = 10.0 M_\odot$ and $R_{\rm c} = 1.00 R_\odot$. From left to right, the columns display the model name, stellar luminosity, Eddington ratio, average mass-loss rate, average asymptotic velocity, average photospheric radius, and the average effective temperature.}             
\label{table: gridresults}      
\centering          
\begin{tabular}{r|cccccc}
\hline\hline       
Model      & $\log_{10}(L_\ast/L_\odot)$ & $L_\ast/L_{\rm edd}$ & $<\dot{M}>$ $[M_\odot/ \rm yr]$ & $<v_\infty>$  $[10^3 \rm km \, s^{-1}]$ & $<R_{\rm ph}>/R_{\rm c}$ & $\log_{10}(<T_{\rm eff}> [\rm K])$ \\
\hline                    
$\Gamma 1$ & $5.33$ & $0.33$ & $1.5 \, 10^{-6}$ & $1.25$ & $1.61$ & $4.99$ \\
$\Gamma 2$ & $5.47$ & $0.46$ & $1.5 \, 10^{-5}$ & $0.62$ & $3.97$ & $4.83$ \\
$\Gamma 3$ & $5.64$ & $0.67$ & $3.2 \, 10^{-5}$ & $1.05$ & $3.64$ & $4.89$ \\
$\Gamma 4$ & $5.74$ & $0.84$ & $7.2 \, 10^{-5}$ & $1.35$ & $4.27$ & $4.88$ \\
\hline                  
\end{tabular}
\end{table*}

\subsection{Radial profiles}

Figure \ref{fig: prof_grid} shows probability density cloud profiles for: i) radial velocity, ii) tangential velocity, iii) gas density, iv) radiation-$\Gamma,$ and v) $\kappa^{\rm OPAL}/\kappa^{\rm line}$-ratio, in the rows for different models with changing boundary luminosity in the columns. 
These probability clouds are computed for each considered quantity by constructing a probability density function from all of the cells in the lateral plane at each given radius. The color-coding in the panels shows the probability density of finding a cell, for example with a given mass density or radial velocity, at a given radial distance from the stellar core (scaled according to $x = 1 - R_{\rm c}/r$ for better visualization). Supplementing these probability clouds, Figs. \ref{fig: 1D_density_prof} and \ref{fig: 1D_velocity_prof} show the average density and radial velocity profiles of all four models $\Gamma1$ - $\Gamma4$. 
Here, we have used empty dots to indicate an estimation of the average sonic point for each profile. This was computed by finding the radius at which the average velocity profile surpasses the average speed of sound of the gas. For all of our models, the average sonic point lies within our simulation domain, where the first few cells close to the bottom boundary are, on average, subsonic. For model $\Gamma 1$, we note that the average wind velocity becomes slightly supersonic after an initial acceleration over the iron opacity bump, before decelerating due to a lack of opacity, and then it reaccelerates again past a new sonic point due to line driving. For this model, only the outermost sonic point has been indicated. Filled dots indicate the mean photospheric radius, for which the computation is explained in Sect. \ref{sec: phorad}.

\subsubsection{Model $\Gamma 3$}
Model $\Gamma 3$ is the 3D base model from the previous section (Sect. \ref{sec: general properties}). 
Here, the material is lifted up directly from layers close to $R_{\rm c}$ by the enhanced opacity associated with the subsurface iron opacity bump. At higher radii, once the density gets low enough, efficient line driving then takes over and ultimately drives the launched wind out of the gravitational well of the star by accelerating it to above the local escape velocity (indicated by dashed red lines in the uppermost panels of Fig. \ref{fig: prof_grid}). Although the range of densities and velocities at each radius is quite large, the average profiles (Figs. \ref{fig: 1D_density_prof} and \ref{fig: 1D_velocity_prof}) are relatively well-behaved. For example, the average radial wind outflow reaches supersonic velocities already close to $R_{\rm c}$, and it continues to rise all the way to the outer boundary. 

\subsubsection{Model $\Gamma 4$}
Model $\Gamma 4$ has a base luminosity higher than the $\Gamma 3$ simulation. The overall wind morphology of this simulation mimics that of the $\Gamma 3$ model, but due to the higher luminosity, it can drive an even higher mass-loss rate out of the stellar potential (see also discussion in Sect. \ref{sec: transition}). We note that for both the $\Gamma 4$ and $\Gamma 3$ simulations, almost all gas parcels have reached velocities above the local escape speed at the outer simulation boundary. This contrasts with the situation in the $\Gamma 2$ simulation, which is discussed next. 

\subsubsection{Model $\Gamma 2$}
In model $\Gamma 2$, as for $\Gamma 3$ and $\Gamma 4$, gas is lifted directly from $R_{\rm c}$ and soon reaches supersonic radial velocities (see the uppermost panel in Fig. \ref{fig: prof_grid}), so that the wind outflow is also initiated by the iron opacity bump below the optical stellar surface. However, due to the lower luminosity, the scale of the radiation force is reduced, which results in the gas particles in the launched wind struggling to reach their local escape speeds. From the probability plot of the radial velocity (uppermost panel in Fig. \ref{fig: prof_grid}), we note that the majority of gas particles have actually not yet reached their escape velocity at the outer simulation boundary, making it uncertain whether they would ultimately make it out of the stellar gravitational well (see also discussion in Sect. \ref{sec: transition}). 
This is also visible in the plot (Fig. \ref{fig: 1D_velocity_prof}) of the average velocity profile, which shows a nonmonotonic behavior in the outer parts where the Rosseland mean opacity does not provide sufficient interaction between the gas and the radiation field, and the effects of line driving start to dominate.
Naively, one might expect a corresponding inversion in the density due to the continuity equation that states that the average $<r^2 \rho v_r>$ is constant throughout the wind. However, in the average density profile corresponding to this model, there is no corresponding inversion in the density. The reasons for this are most likely related to:  i) $\langle \rho \varv \rangle \ne \langle \rho \rangle \langle \varv \rangle,$ and ii) a stronger than $r^{-2}$ decline in density.
Essentially, this $\Gamma 2$ simulation has initiated a very dense outflow directly from the iron opacity bump, but due to its reduced luminosity, the launched mass flux cannot be efficiently carried outward, and so some of the dense gas filaments begin to stagnate and might even fall back upon the core. This can also be directly seen from the probability plots, where this simulation not only shows a very large range of velocities but also displays larger downward motions than the other simulations, reaching many hundreds of kilometers per second. 

\subsubsection{Model $\Gamma 1$}
In model $\Gamma 1$, the bottom boundary luminosity is even lower. This simulation shows a markedly different behavior than those with higher base luminosities. In other words, in this model, the opacity around the iron opacity bump is not enough to launch a radial wind outflow from these deep subsurface regions. Instead, a very turbulent, extended atmosphere is created, which does not experience an outward-directed supersonic average velocity. In other words, although the velocity dispersion is still highly supersonic, a net radial wind outflow is no longer launched. 

What we obtain here is a very extended and turbulent atmosphere, reaching $x \approx 0.4, r \approx 1.67$ where the density has become low enough for line driving to take over and launch a (significantly less dense, see also below) supersonic wind. In the deep surface regions, the opacities in this model correspond to the iron opacity bump at $\sim 200 \, kK$ rather than the opacity increase due to Helium \citep{Grassitelli2018}. Helium recombination dominates the Rosseland mean at $\sim 50 \, kK$, which in this model occurs at $x \approx 0.7$. However, in those regions, the gas has already reaccelerated due to line driving, which is the main contributor to the total opacity in the outer wind (see the bottom left panel in Fig. \ref{fig: prof_grid}).

\subsection{Photospheric radius} \label{sec: phorad}
Using the hydrodynamical structure and the OPAL Rosseland opacity tables, a photospheric radius $R_{\rm ph} (\tau_{\rm Ross} = 2/3)$ can be derived from the models. For simplicity, this is done by computing the Rosseland mean optical depth in the radial direction. Since the models are time-dependent and not spherically symmetric, the photospheric radius will also vary with both time and lateral position. Figure \ref{fig: R_ph_r} shows the distribution of photospheric radii encountered in ten different snapshots at all different lateral points for the $\Gamma 3$ model. From this distribution, a mean photospheric radius $<R_{\rm ph}>$ is computed by taking the straight average, which can then be compared to the radius of the hydrostatic core $R_{\rm c}$. 

Table \ref{table: gridresults} lists the mean photospheric radii for the four 3D models. There is a monotonic increase in the photospheric radius with the Eddington ratio. Here the $\Gamma 1$ model stands out, having a significantly smaller photospheric radius than the $\Gamma2 - \Gamma4$ models. While the $\Gamma 2-\Gamma 4$ models have very optically thick winds with $\langle R_{\rm ph} \rangle/R_{\rm c} \approx 4$, $x \approx 0.75$, for the $\Gamma 1$ model $\langle R_{\rm ph} \rangle/R_{\rm c} \approx 1.6$, corresponding to $x \approx 0.4$.

\subsection{Effective temperature}
\begin{figure}[h]
\centering
\includegraphics[width=0.5\textwidth]{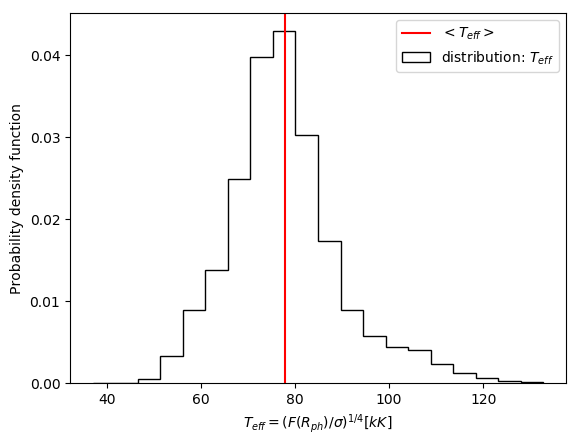}
  \caption{Probability density function of the effective photospheric temperature for the $\Gamma 3$ model. The horizontal axis shows the effective temperature for a particular radial line of sight. Due to the time-dependent and laterally structured wind, this varies for different radial lines of sight, giving a distribution in $r$. This distribution is computed based on $128 \times 128$ transverse cells and on $10$ time snapshots. The effective photospheric temperature is defined here as a function of the radiation flux calculated at the photospheric radius. The red line indicates the mean of the distribution. The probability density function is normalized such that the integral over the variable shown on the x-axis gives unity.}
     \label{fig: T_eff}
\end{figure}

Continuing this analysis, a corresponding effective temperature $\sigma T_{\rm eff}^4 \equiv F (R_{\rm ph})$ can be computed, where $F (R_{\rm ph})$ is the radiative flux at the photosphere and $\sigma$ is the Boltzmann constant. At every lateral point, the radial component of the FLD-flux is taken, using Eq. \eqref{eq: fld_flux} at the photospheric radius. Again, due to the variation in time and lateral space, there is a distribution in effective temperatures. This is shown in Fig. \ref{fig: T_eff} for the $\Gamma3$ model.

For each of the four 3D models, the average effective temperature is computed and listed in Table \ref{table: gridresults}. Models $\Gamma2$-$\Gamma4$ have comparable effective temperatures. Here, two effects cancel each other out. Model $\Gamma4$ has a higher Eddington ratio and thus a higher stellar luminosity than model $\Gamma2$, but its photospheric radius is further away from the hydrostatic core, and the radiative flux drops, on average, by $1/r^2$. On the other hand, model $\Gamma 2$ has a lower stellar luminosity, but the flux is evaluated closer to the hydrostatic core due to the smaller photospheric radius. Apart from this, model $\Gamma1$ stands out again with a significantly larger effective temperature. This also indicates different behavior in the overall wind structure.

\begin{figure}[h]
\centering
\includegraphics[width=0.5\textwidth]{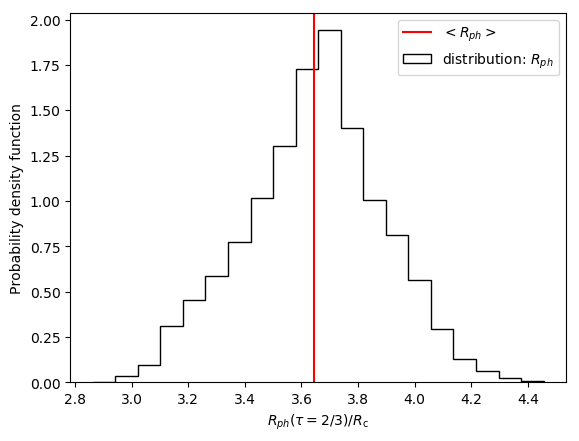}
  \caption{Probability density function of the photospheric radius for the $\Gamma 3$ model. The horizontal axis shows at the radius at which the radial Rosseland mean optical depth is equal to $2/3$. Due to the time-dependent and laterally structured wind, this varies for different radial lines of sight, giving a distribution in $r$. This distribution is computed based on $128 \times 128$ transverse cells and on ten time snapshots. The photosphere is defined here as the radius where the Rosseland mean optical depth is equal to 2/3. The red line indicates the mean of the distribution. The probability density function is normalized such that the integral over the variable shown on the x-axis gives unity.}
     \label{fig: R_ph_r}
\end{figure}

The histograms in Figs. \ref{fig: R_ph_r} and \ref{fig: T_eff} further display the simulated variation of the photospheric radius and effective temperature in the $\Gamma 3$ model, showing a full width half maximum of about $20$ kK in $T_{\rm eff}$ and $0.5\, R_{\rm c}$ in photospheric radius. 
These variations reflect the fact that the $\tau_{\rm Ross} = 2/3$ surface along a radial line through the fast, low-density lanes of the simulations (e.g., Fig. \ref{fig: timeseries_rho_v}) lies much deeper than where such a radial line hits high-density clumps in the outer wind layers. 

We caution, however, that we have used a very simple radial integration to define a photospheric radius and the corresponding effective temperature along each line of sight (see above). The width of the distribution thus serves only as a first visualization of the variability, and will likely not be representative of the actual variation in stellar effective temperature. In follow-up work, we will reinvestigate this by means of 3D radiative transfer \citep{Hennicker2020, Hennicker2022} performed on the full wind volume. This will allow us to investigate the typical level of photometric and line-profile variability that our 3D models would predict. 

\subsection{Mass-loss rate} \label{sec: mdot}
\begin{figure}[h]
\centering
\includegraphics[width=0.5\textwidth]{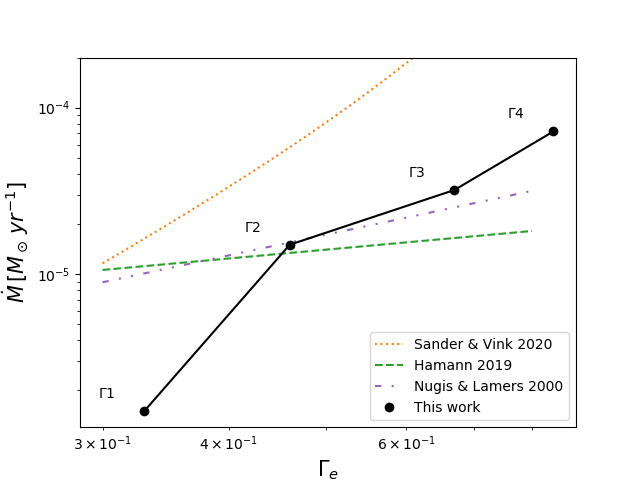}
  \caption{Relation between mass-loss rate and the Eddington factor over the grid of four 3D wind models. The models are compared to empirical data from \citet{Hamann2019} (dashed green  line) and \citet{NugisLamers2000} (dash-dotted purple  line), and theoretical models from \citet{Sander2020} (dotted orange line). }
     \label{fig: Mdot_gam}
\end{figure}

Furthermore, we also compute average mass-loss rates from our models, by taking a time and spatial average of the mass-loss rate $\dot{M} = 4 \pi \rho v_r r^2$ flowing out from the outer boundary of the simulation box.   
Fig. \ref{fig: Mdot_gam} compares the average mass-loss rate $\langle \dot{M} \rangle$ with various results for steady WR-type mass-loss rates from the literature. This comparison shows fair agreement in the WR wind regime ($\Gamma2$-$\Gamma4$) with the empirical results by \citet{NugisLamers2000} and \citet{Hamann2019}. For the $\Gamma1$ simulation, our $\langle \dot{M} \rangle$ is significantly lower than these empirical scalings, which stem from the analysis of dense WR outflows. Our rates are also considerably lower than the recent 1D steady-state models by \citet{Sander2020}. The reason for this discrepancy may be  that, in their 1D stationary models, \citet{Sander2020}  needed to (somewhat artificially) boost the radiative acceleration to enforce a monotonic velocity field, ensuring that the full mass flux initiated at the sonic point escapes. They did this by invoking an ad hoc large amount of clumping in the models ($f_{\rm cl} =50$, see definition of $f_{\rm cl}$ below), assuming that all wind mass is contained within these high-density clumps, albeit while still solving the steady-state equation-of-motion for the smooth (rather than clumped) wind. 

\begin{figure*}[ht]
\centering
\includegraphics[width=\linewidth]{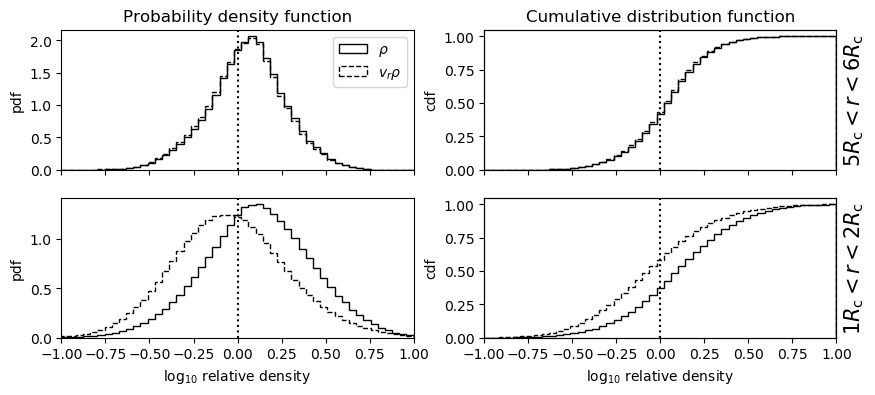}
  \caption{Distributions of mass (solid lines) and radial momentum (dotted lines) in our $\Gamma 3$ model as a function of relative density. The two panels on the left show the probability density function, and the two panels on the right show the cumulative distribution function. The probability density and cumulative distribution functions calculated for the regions in the wind between $5 R_{\rm c}$ and $6 R_{\rm c}$, and  between $1 R_{\rm c}$ and $2 R_{\rm c}$, are shown in the top and bottom panels, respectively. The probability density functions are normalized such that the integral over the variable shown on the x-axis gives unity.} 
     \label{fig: mass_dist}
\end{figure*}

The natural question then arises about the fraction of the wind mass contained within the high-density parts of our simulations. Figure \ref{fig: mass_dist} shows the distribution of mass and radial momentum as a function of relative density for our $\Gamma 3$ model. The top panels display the mass and outward momentum distribution from the lowermost region of the simulation, between $5 R_{\rm c}$ and $6 R_{\rm c}$, and the bottom panels display the distributions between $1 R_{\rm c}$ and $2 R_{\rm c}$. Here, the probability density functions are computed by first binning the cells according to their relative density. Then, for each bin in relative density, the number of cells with that corresponding relative density are counted and weighted with their density or momentum. Afterward, the probability densities are normalized such that the integral of the probability density function over relative density gives unity. 

In both the upper and lower panels, it is seen that a little over half of the mass is located in the overdense regions, to the right of the vertical dotted line indicating the average density. From the bottom panel, one can also see that, close to the stellar core, a majority of the momentum is contained in the high-velocity, low-density material. However, further out in the wind (upper panels), a majority of the momentum, and thus mass-loss rate, is contained within gas with a higher-than-average density. Most notably, however, the overall distributions of mass and momentum in our simulations are actually quite broad and Gaussian-like, which is very different from the typical two-component medium (clumps and an inter-clump medium) ansatz assumed in various spectroscopic and diagnostic clumpy wind models (e.g., \citealt{Hillier1998,SundqvistPuls2018,Sander2018}). 

\subsection{Transitioning from an optically thick to an optically thin wind } \label{sec: transition}

By varying the input stellar luminosity, the models presented here show a natural, smooth transition over different wind morphologies. At the highest luminosities, simulations $\Gamma3$ and $\Gamma4$ essentially represent the winds from classical WR stars, where the material is lifted up directly from layers close to $R_{\rm c}$ by the enhanced opacity associated with the subsurface iron opacity bump. With lower luminosity,  model $\Gamma 2$ is quite similar to the dynamically inflated 1D models presented in \citet{Poniatowski2021}, where an ad hoc increase in the line force in the outer regions had to be made by the authors so that the wind launched at the iron opacity bump could reach a velocity above the local escape speed. At the lowest simulated luminosity, model $\Gamma 1$ resembles how a standard line-driven wind is launched from the optically thin parts of the atmospheres in, for example, main-sequence O-stars (e.g., \citealt{Bjorklund2020}), rather than from the subsurface iron opacity peak. The main difference between such a main-sequence O-star and the simulation here is that here the star is much more compact (meaning that it has a much smaller radius). Moreover, the subsonic average structure and the substantially reduced mass-loss rate observed in our $\Gamma 1$ simulation are in qualitative agreement with the 1D (stationary and subsonic) hydrodynamic stellar structure models by \citet{Grassitelli2018}, who found that below a certain mass-loss limit, a classical WR star is not able to launch a supersonic wind from the iron opacity bump region. This is apparent in Fig. \ref{fig: 1D_velocity_prof}, where the location of the average sonic point for the different models has been indicated with an empty circle
(see also \citealt{Sanyal2015}). However, in our $\Gamma 1$ simulation, the driving force behind the second acceleration through the sonic point is different. In \citet{Grassitelli2018}, the reacceleration occurs due to a second bump in the Rosseland mean opacity (e.g., the helium opacity bump), while in our models it is mainly a line-driven wind that is optically thin in the continuum. The extent of the inflated and optically thick turbulent atmosphere in our model is quite large, covering $\sim 1.61 \, R_\odot$ in distance above the core radius. This model has a mass-loss rate that is, for the same stellar mass, almost an order of magnitude lower than the simulations, with a higher base Eddington factor.
Overall, this $\Gamma 1$ model is probably a quite good representation of the transition from WR stars to the stripped hot (sub)dwarfs that are believed to be results of binary evolution \citep{Han2010}, and it essentially represents low-luminosity counterparts of classical WR stars \citep{Gotberg2018}. 


Thus, our grid of simulations with varying luminosity demonstrates the transition from a moderately inflated and very turbulent atmosphere with a "standard" optically thin line-driven wind on top of it ($\Gamma 1$), to a classical WR star with a very optically thick and supersonic wind outflow driven directly from deep subsurface regions (>$\Gamma 2$). This transition is also directly reflected in the significant drop in the average mass-loss rate observed for the $\Gamma 1$ simulation, as discussed below (see also Fig. \ref{fig: Mdot_gam} and Table \ref{table: gridresults}). 
Moreover, although all of our simulations are still accelerating at the simulated outer boundary, from the slope of the average velocity curves in Fig. \ref{fig: 1D_velocity_prof}, we may deduce that the $\Gamma$ 1 simulation likely would reach a final terminal wind speed that is significantly higher than the other models (consistent also with the 1D, stationary models by \citealt{Sander2020}). 

The transition from optically thick to optically thin wind conditions is also directly reflected in the average photospheric radius $\langle R_{\rm ph} \rangle$ of the $\Gamma 1$ model as compared to the simulations with higher luminosity (see Table \ref{table: gridresults}). 
%
Since the supersonic wind in this simulation is launched only at $x \approx 0.4$ (see Fig. \ref{fig: 1D_velocity_prof}), the location of $R_{\rm c}$ verifies that the supersonic radial outflow in the $\Gamma 1$ model indeed represents a quite standard optically thin line-driven wind initiated from layers above, but still quite close to the optical stellar surface. 


In this respect, we note that although the mass flux through the outer simulation boundary in the transitional $\Gamma 2$ simulation suggests a high WR-type mass-loss rate for this model, it is not entirely clear that this would be the final stellar mass-loss rate. In other words, although the gas parcels at the outer boundary of this simulation have positive radial velocities, the majority of them have still not reached their local escape speeds (see Fig. \ref{fig: prof_grid}). As such, their final fates are still somewhat uncertain; it is unknown whether they will eventually be able to escape the stellar potential, or instead lose their outward
 momentum and ultimately start falling back upon the core. In order to investigate this, the current simulations would have to be extended to higher radii, which we defer to future work. It is nonetheless interesting to note that the nature of these transitional winds may be somewhat different from their counterparts with higher (classical WR winds) and lower (hot subdwarfs) luminosities. Because of the high mass flux initiated in the subsurface regions, the transitional wind will still be optically thick, and as such it resembles that of a classical WR star. However, in comparison to stars with higher luminosities, the high-density regions in this wind (likely the most visible parts in spectral observations) should be characterized by significantly lower radial wind speeds.

\subsection{Clumping factor and velocity dispersion} \label{subsec: clumping}

1D stellar atmosphere codes such as PoWR \citep{Grafener2002}, {\sc cmfgen} \citep{Hillier1998} and {\sc fastwind} \citep{Puls2020} typically assume an analytic expression for the velocity and density profile, and thus they do not directly model structure in the wind. Instead, such codes rely on different parameterizations of the complex structure in the wind. Our 3D models now allow us to gauge these parameterizations. In this section, we look at the clumping factor and the velocity dispersion, which are both important quantities when deriving stellar and wind parameters from observed spectra.

\subsubsection{Clumping factors} 
The clumping factor is important in the context of spectral diagnostics, for example when investigating recombination lines to determine mass-loss rates. To characterize overdensities, the clumping factor is defined as

\begin{equation}
    f_{\rm cl} \equiv \frac{\langle \rho^2 \rangle}{\langle \rho \rangle^2},   
\end{equation}
where the angle brackets here denote averaging over time and lateral directions. Since the absorption coefficients (per unit length) of optical recombination lines scale as $\chi \sim \rho^2$, the mean absorption becomes enhanced in a clumped outflow. 
In a stellar wind, the mean mass-loss rate scales with the density: $\langle \dot{M} \rangle \sim \langle \rho \rangle$, so for the absorption coefficient of a recombination line: $\langle \chi \rangle \sim \langle \rho^2 \rangle \sim f_{\rm cl} \langle \rho \rangle^2 \sim f_{\rm cl} \langle \dot{M} \rangle^2$. It is well known that this can have a significant effect upon empirical mass-loss rate determinations in hot star winds (see \citealt{Puls2008}, for a review). 

\begin{figure}[h]
   \centering
   \includegraphics[width=0.5\textwidth]{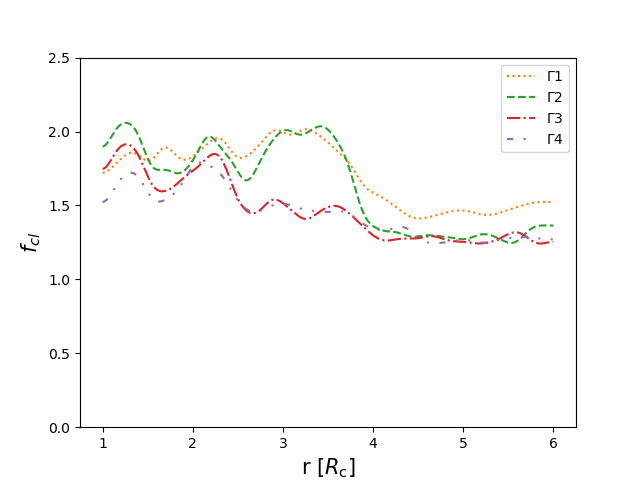}
      \caption{Clumping factor as a function of radius for each of the four 3D models in our grid.}
         \label{fig: clumping}
   \end{figure}

Figure \ref{fig: clumping} shows that the clumping factor is on the order of $f_{\rm cl} \sim 2$ in our 3D simulations. We note that this is significantly lower than seen in corresponding 2D models, where it rather lies between $\sim 5-10$. Again, this reflects the fact that density contrasts in general are sharper in 2D models than in 3D models, because in 3D, gas parcels have one more dimension in which they can spread out, leading to a smoothing effect (see discussion in Sect. \ref{sec: 2Dv3D}). 


As can be seen from Fig. \ref{fig: lat_slice}, the physical scales of the clumps vary throughout the radial extent of the simulation. By inspecting and measuring the diameter of some of the typical over-dense regions, we can make some simple first estimates of the characteristic clump sizes. In all of the simulations, we find that the typical lateral length scale of the over-densities is about $\sim 0.02 R_{\rm c}$ at a radius of $r = 1 R_{\rm c}$, $\sim 0.06 R_{\rm c}$ at a radius of $r = 2 R_{\rm c}$, and roughly constant at $\sim 0.1 R_{\rm c}$ at radii outward of $r = 3 R_{\rm c}$. 

\subsubsection{Velocity dispersion} \label{sec: vdisp}
\begin{figure}[h]
   \centering
   \includegraphics[width=0.5\textwidth]{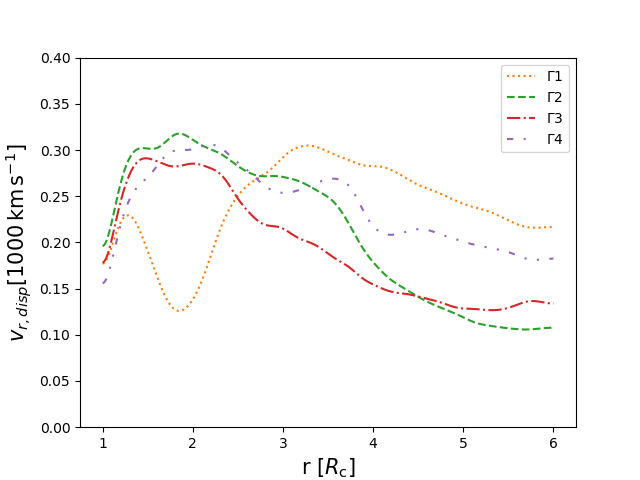}
      \caption{Radial velocity dispersion as a function of radius for each of the four 3D models in our grid.}
         \label{fig: vdispersion}
   \end{figure}

\begin{figure}[h]
   \centering
   \includegraphics[width=0.5\textwidth]{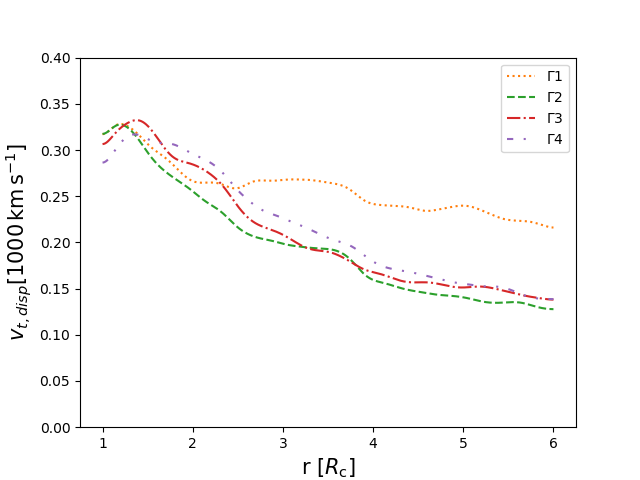}
      \caption{Tangential RMS velocity for each of the four 3D models in our grid.}
         \label{fig: vtdispersion}
   \end{figure}

The probability density maps discussed above directly suggest large velocity dispersion. We distinguish here between a radial component: 

\begin{equation}
    v_{r, \rm disp} = \sqrt{ <v_r^2> - <v_r>^2 } 
\end{equation}
%
and a lateral component. Since there is no net displacement in the lateral direction, the average lateral component of the velocity vector should be zero. However, this does not mean that the average magnitude of the velocity vector projected on the lateral plane is zero. As such, to statistically quantify lateral motions, we here inspect directly the root mean square (RMS) of the projected velocity vector: 

\begin{equation}
    v_{t, \rm RMS} = \sqrt{  \langle v_t^2 \rangle }, 
\end{equation}
where $v_t^2 = v_x^2 + v_y^2$. Figures \ref{fig: vdispersion} and  \ref{fig: vtdispersion} show the time and laterally averaged radial velocity dispersion, and the lateral RMS velocity of the tangential velocity, as a function of radius. Models $\Gamma2-\Gamma4$ have a peak in their radial velocity dispersion at around $2\, R_{\rm c}$, after which the value steadily declines in the outer wind. The $\Gamma1$ model, on the other hand, peaks significantly further out, at around $3.5\, R_{\rm c}$. This aligns well with the interpretation that, for the more luminous models, the static Rosseland mean opacity only barely suffices to drive the stellar gas that is launched close to $R_{\rm c}$ outward, resulting in coexisting regions of up- and down-flowing material close to the stellar core. However, when line driving takes over further out in the wind, material can escape much more easily. For the less luminous model $\Gamma1$, on the other hand, the wind is only initiated at $\sim 1.5\, R_{\rm c}$ (see discussion above), which means that the vigorous turbulence initiated around the iron opacity peak actually starts to decrease somewhat before the line-driven wind is ignited. As seen from both the probability-density map of the radial velocity (upper left panel in Fig. \ref{fig: prof_grid}) and the radial velocity dispersion as a function of radius (Fig. \ref{fig: vdispersion}), this then gives a distinctly different behavior, characterized by large dispersion close to the core radius, followed by first a region of declining values and then a second rise when the line-driven wind is launched. 

Typical upper values for both radial and tangential quantities in all four models are found to be $\sim 300 \, \rm km/s$. For the higher luminosity models with WR-like mass loss, this means that when the background radial velocity field is subtracted, the turbulent motions are close to isotropic. However, as discussed in Sect. \ref{sec: mod lim}, at least for the outer wind (which experiences significant line driving), this velocity-dispersion isotropy needs to be further investigated and anisotropic line opacities need to be accounted for.  

\section{Discussion} \label{sec: Discussion}
The successful 3D RHD simulations presented in this paper overall provide precious information about the structure and characteristics of WR stellar atmospheres and wind outflows. In addition, they suggest a natural transition between the dense, optically thick outflows from high-luminosity WR stars and more standard optically thin line-driven wind outflows from hot \mbox{(sub)dwarfs}. 
Nonetheless, there are a number of simplifications we have made in order to make our 3D computations possible, in particular regarding the calculation of the radiative acceleration. 

\subsection{Model limitations} \label{sec: mod lim}
First, our tabulations of the line-force parameters are based on calculations assuming the radiation and gas temperatures are equal. Although this approximation has been shown to provide good estimates of the line force also in the O-star domain \citep{Poniatowski2022}, it is simultaneously well known that, at least when performing spectral synthesis, the outer regions of WR outflows are affected by NLTE effects. As such, it would be useful to extend our current tabulations toward (approximate) NLTE conditions (following, e.g., \citealt{Puls2000}), in order to examine how this might affect the tabulations of the line-opacity parameters. 

Secondly, in this paper we have assumed an isotropic $\kappa^{\rm line}$ set by the radial velocity gradient; since the examined WR flows are strongly dominated by the radial velocity component in the regions with significant $\kappa^{\rm line}$, this is probably a quite reasonable first approximation. However, a more realistic treatment would instead use an anisotropic $\kappa^{\rm line}$ computed from a proper average of the velocity gradients along different directions. Such considerations of nonradial velocity gradients might then affect, for example, the tangential velocity dispersion in the outer wind (see Sect. \ref{sec: vdisp}). Moreover, 
the radial line acceleration might also be affected by nonradial velocity gradients, due to the solid angle integration that is formally required in the computation of  this component as well \citep[e.g.,][]{Pauldrach1986}. 

Thirdly, our computation for $\kappa^{\rm line}$ is essentially carried out using the Sobolev approximation. This neglects the strong line-deshadowing instability (LDI) inherent to line-driven flows \citep{Owocki1988}. Although this LDI causes strong levels of wind clumping in OB-star winds (e.g., \citealt{Driessen2019}), linear analysis suggests that it may be damped within dense WR outflows \citep{Gayley1995}. Indeed, empirical analysis suggests lower levels of clumping in WR outflows \citep{Hillier1991} than in O-stars \citep{Hawcroft2021}. Nonetheless, as shown in Fig.  \ref{fig: clumping}, the clumping factors found in these first 3D simulations, $f_{\rm cl} \sim 2$, are certainly on the lower end of what seems to be suggested by analysis of electron scattering wings in combination with optical recombination lines (see overview in \citealt{Crowther2007}). This suggests that the strong radial compression typically induced by the LDI may still play a role in setting the absolute scale of clumping also in WR outflows, at least in the outer wind parts with modest continuum optical depths. 

Next, our setup is that of a Cartesian box-in-wind simulation, where we correct for the $1/r^2$ decline of the fluxes of the conserved quantities in PDEs \eqref{eq: hd_rho}-\eqref{eq: hd_e} and \eqref{eq: rhd_E} (see also appendix A in \citealt{Moens2021}). However, so far it is unclear how wind structures would deform and grow or diffuse when advected outward in a lateral direction due to the divergence of the radial directions at different points on a lateral plane. Taking this effect into account would require moving away from our current "pseudo-planar" approach, using either one of two methods. A first possibility is to perform global simulations in a "wind-in-box" (instead of box-in--wind) setup. The drawback with this method is that the resolution, by necessity, would be lower, and thus it is unclear how well structures would be resolved (especially in the deep, critical layers around the mean sonic point). Alternatively, one could perform local box-in-wind simulations as presented here, but in full spherical coordinates. This would not require significantly higher resolution, but is so far not possible within our numerical setup due to the spherical divergence operator that would be present in Eq. \eqref{eq: rhd_E}.

\subsection{Comparison between hybrid-opacity flux-limited diffusion and short characteristics}
\begin{figure}[h]
   \centering
   \includegraphics[width=0.5\textwidth]{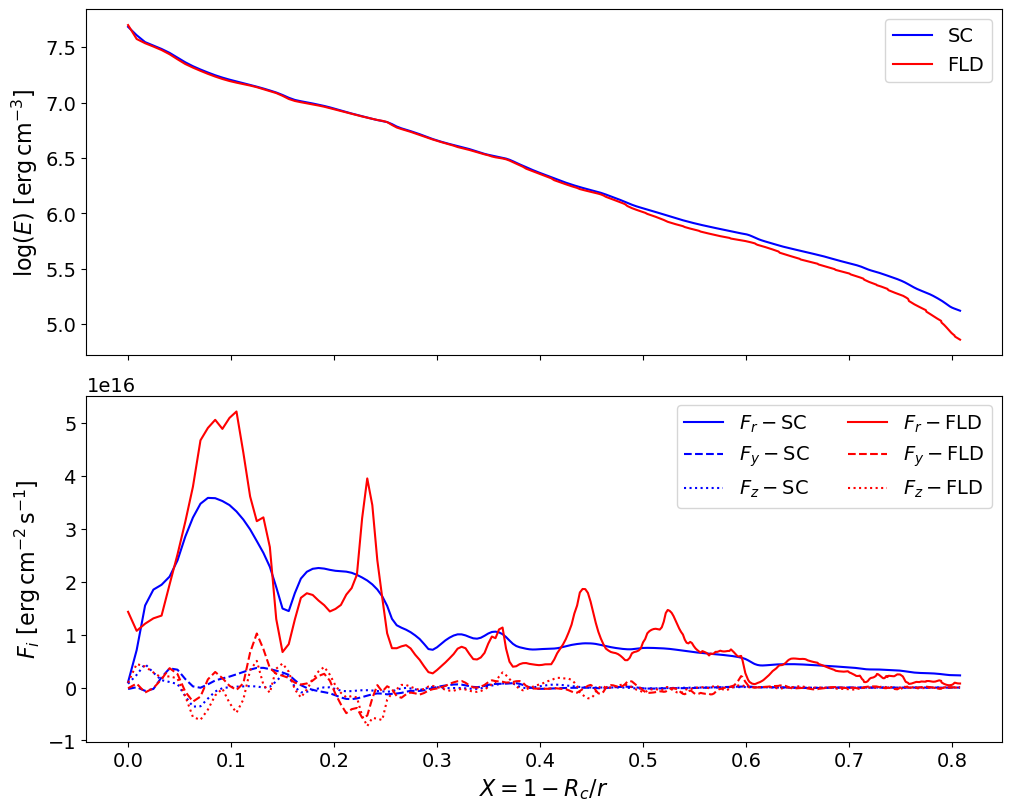}
      \caption{Radial stratification in the lateral midplane (y=z=0) of the radiative energy density (top panel) and the flux components (bottom panel) as obtained for a representative snapshot of the 3D $\Gamma 3$ simulations. The radiation quantities are calculated from the FLD method (red lines) or from a detailed 3D radiative transfer solution technique using short characteristics (blue lines).}
         \label{fig:sc_fld}
   \end{figure}

Finally, our flux calculations rely on analytic closure relations between radiation energy density, pressure, and flux.
To estimate the error introduced by these analytic relations used in our FLD approach, we have recalculated the radiation quantities with a full 3D short characteristics (SC) solution method for a representative simulation snapshot (corresponding to Fig.~\ref{fig: lat_slice}). Within the SC method, we are solving the (gray) equation of radiative transfer for the specific intensity directly, and then perform the angular integrals to obtain the radiation energy density and radiation flux components. To this end, we have extended the SC radiative transfer solution scheme from \cite{Hennicker2020} to include periodic boundary conditions in a slab geometry.
Since this solution framework requires a global simulation setup (in contrast to the local FLD approach), we remap all required quantities on a global spatial grid with $N_r=301$ grid points using a logarithmic spacing to recover a high resolution near the stellar core, and an equidistant grid spacing with $N_y=N_z=101$ in the lateral components. The angular integrals are then performed by using $N_\Omega=32$ directions distributed symmetrically over all octants. Fig.~\ref{fig:sc_fld} shows the resulting radial stratification of the radiative energy density and flux components for the 3D $\Gamma 3$ model, compared to the corresponding solution from the FLD method.

Particularly for the energy density, the differences between the two frameworks are only marginal. In the outer wind, we do find some deviations, which, however, are most likely related to the SC method. Namely, within the slab-geometry SC scheme, the finite size of the stellar atmosphere is intrinsically neglected by using periodic boundary conditions. Moreover, since at present it is somewhat unclear how to best include line opacity in the SC scheme, these calculations have been carried out for a total $\kappa = \kappa^{\rm OPAL}$. However, while $\kappa^{\rm OPAL}$ dominates the contribution in the inner atmospheric parts, in the outer parts, line opacity can also play a role  in the scale of radiation energy density (see previous discussions). 
The flux components show somewhat larger quantitative differences between the FLD and SC solution schemes, with a smoother stratification found from the SC method. The qualitative behavior, however, is very similar. In particular, considering the overall scale, for each of the flux components there is a good agreement between the two methods, with a much higher radial flux
when compared to the lateral components, as  also found from the FLD solution. 
Overall, the qualitative behavior of all radiation quantities is similar, lending support to the analytic closure relations used in this paper.  

\section{Summary and future work} \label{sec: Conclusions}
In this paper, we describe the first successful 2D and 3D numerical models of the radiation-driven atmospheres and outflows from WR stars. Our simulations have been calculated using a box-in-wind approach, in which the equations of hydrodynamics are solved together with the frequency-integrated first moment of the radiative transfer equation using an analytic flux-limiting closure relation, relating the time-dependent radiative energy densities and fluxes. In this way, we are able to account for the time-dependent radiation force, as well as heating and cooling source terms.  Opacities are described using our hybrid approach that combines Rosseland mean opacities in the static limit with a description of velocity-stretched line-driving opacities, computed from an atomic database consisting of some four million spectral lines \citep{Pauldrach1998, Pauldrach2001}.


In our multidimensional models, we find that gas close to the stellar core is accelerated by a radiation force that is mainly interacting via the Rosseland mean opacity. 
However, not all gas is immediately driven outward. Due to a combination of the strong radiation force and  instabilities associated with subsurface opacity peaks, gas clumps together creating turbulent motions and the coexistence of highly supersonic up- and down-flowing gas. We note that this structure formation is not a result of the LDI, such as is present in OB-stars \citep{Driessen2019}, but most likely arises due to convective instabilities associated with subsurface opacity peaks in high-luminosity stars near the Eddington limit (\citealt{Jiang2018}, see also \cite{Castor2004}, their Ch. 7.3).
The gas that reaches far enough away from the (quasi)hydrostatic core is then further accelerated by line driving. However, clumps that are further out in the wind are too dense to experience efficient line driving, and continue to advect outward via ram pressure from the faster, line-driven low-density gas. It is really this interplay between radiation driving on low-density gas and ram pressure from low-density gas acting on high-density gas that finally allows the material to escape the stellar potential.

After identifying this fundamental mechanism, we studied the influence of the core luminosity. For this, a first grid of 3D models was calculated, where the bottom boundary input luminosity was varied. We found a transition from the strong optically thick wind as observed in classical WR stars to an optically thin line-driven wind of a hot, compact subdwarf star.
This transition shows itself in the mass-loss rates, terminal velocities, location of the mean photospheric radius, and the sonic point location (see also the corresponding predictions from the 1D, stationary models by \citealt{Grassitelli2018, Sander2020}).
For each of the four 3D models, we characterized mean velocity and density profiles. Just by lowering the bottom luminosity, the wind launching in the hot subdwarf is delayed until $\approx 1.5\, R_{\rm c}$. Below this launching radius, the atmosphere is 
very structured and turbulent, but without a significant net radial velocity. Hereafter, a line-driven wind launches with a density that is an order of magnitude lower than the WR models, resulting in an optically thin wind with a significantly lower average mass-loss rate. 
In our simulations, the transition point between the different types of winds happens at a threshold around $L_\ast/L_{\rm Edd} \lesssim 0.4$ (see Table \ref{table: gridresults}). Stars with an Eddington ratio lower than this threshold would thus appear as hot subdwarfs featuring an optically thin wind, while stars above it would appear as WR stars featuring a thick wind. We note, however, that this limit will also depend on stellar metallicity (which influences the opacities) and rotation (which may reduce the effective gravity); effects from this will be investigated in a planned follow-up work. 

Additionally, the 3D models were used to compute clumping factors and velocity dispersions, which are on the order of $f_{\rm cl}\approx 2$ and $v_{\rm r, \rm disp} \approx v_{\rm t, \rm RMS} \approx 100-300\,  \rm km \, \rm s^{-1}$ for all four models. Concerning the clumping factors, this is significantly lower than what is found in our corresponding 2D models, which feature sharper gas density structures, suggesting that it is indeed important to model these types of outflows in 3D rather than in 2D. As discussed in Sect. \ref{sec: mod lim}, the $f_{\rm cl} \approx 2$ found here lies at the lower end of what is typically inferred from observations of electron scattering wings, suggesting that the strong radial compression associated with the LDI may ultimately play a role in setting the absolute scale of wind clumping also in WR stars (in combination with the subsurface, convection-driven instabilities and structures investigated here). 
Average mass-loss rates calculated from our WR-like models further agree reasonably well with trends observed in empirical studies of WR stars \citep{NugisLamers2000,Hamann2019}, while the predicted mass-loss rate of the hot subdwarf model is about an order of magnitude lower.

A natural follow-up on the work presented here is the analysis of synthetic light curves and spectra computed from our 3D models. WR stars have been observed to be prone to small substructures in spectral lines, so-called line-profile variations (LPV's), due to high-density clumps moving through the wind \citep{Lepin1996, Lepin1999}. Our models allow us to compute synthetic spectral lines from consecutive snapshots and, in that way, study how our predicted structures compare to observed LPVs in WR stars. In the same spirit, the variation of the optical photosphere may also result in stochastic variations of photometric light curves, as indeed observed for WR stars \citep[e.g.,][]{Lenoir2022}. Moreover, \citet{Chene2020} suggested that the specific behavior of LPV and clumping observed for different WR stars indeed might indicate that their wind structure predominately arises because of convective instabilities, rather than the LDI (see also suggestions in \citealt{Grassitelli2016}). 


In future work, two of the simplifications mentioned in Sect. \ref{sec: Discussion} can be addressed within our setup. The first one is the assumption of an isotropic line opacity. This can be circumvented by reformulating the FLD closure with a nonisotropic diffusion coefficient, allowing us to more accurately simulate lateral radiation forces. The second is the box-in-wind setup, which may be improved upon by restructuring our simulations in a global, Cartesian star-in-box model. This might allow us to more accurately follow the wind structures further away from the star, although at the cost of a larger simulation space with an overall lower resolution. In addition, this would allow
 us to study the influence of rapid stellar rotation on the wind structure.

Furthermore, the methods used in this work are general in such a way that they can be expanded toward other regions of the Hertzsprung-Russel diagram. So far we have covered outflows from classical WR stars and a single model of a hot subdwarf, but similar methods could be used to study, for instance, the interaction between a turbulent atmosphere and the overlying line-driven wind in OB-stars.
Additionally, methods described in this work could prove valuable in the modeling of wind-envelope interactions in S-Dor type variables \citep{Grassitelli2021},
and the influence from line driving on these stars' turbulent envelopes and outflows \citep{Jiang2018}. In these studies, first efforts have been made in understanding the variability of these stars as a consequence of iron and helium bump opacities, but previous models have failed to reproduce expected mass-loss rates. In view of our results here, it seems likely that this might be due to the negligence of line driving. 
In general, the combination of FLD-RHD with our hybrid opacity prescription indeed has potential for a quite broad array of topics. 

\begin{acknowledgements}
      The computational resources and services used in this work were provided by the VSC (Flemish Supercomputer Center), funded by the Research Foundation Flanders (FWO) and the Flemish Government – department EWI.
      NM, LH, and JS acknowledge support by the Belgian Research Foundation Flanders (FWO) Odysseus program under grant number G0H9218N. 
      LP and JS acknowledge support from the KU Leuven C1 grant MAESTRO C16/17/007.
      IEM has received funding from the European Research Council (ERC) under the European Union’s Horizon 2020 research and innovation programme (grant agreement No 863412)
      The National Solar Observatory (NSO) is operated by the Asociation of Universities for Research in Astronomy, Inc. (AURA), under cooperative agreement with the National Science Foundation.
      The authors are grateful to the referee, Luca Grassitelli, for his constructive input to the manuscript.
\end{acknowledgements}

%
%

\bibliographystyle{aa}

\bibliography{WR_winds}

\newpage

\begin{appendix}


\section{Convective instability with radiation pressure}\label{App: BruntVaisala}
Following  \citet{Castor2004} (their Ch. 7), we here examine further the instability that arises in our simulations. 
Namely, the atmospheres of luminous stars are prone to instabilities when the Eddington factor approaches unity ($\Gamma \approx 1$). \cite{Castor2004} performed a linear stability analysis on the momentum equation \eqref{eq: hd_mom}, stating that the atmosphere becomes absolutely unstable where the radiation-modified Brunt-V\"ais\"al\"a frequency $\omega^2_{\rm BV}$ becomes imaginary:

\begin{equation}
    \omega^2_{\rm BV} = \frac{(\gamma -1)a^2}{\gamma^2 H^2}\left[1-(n+q) \frac{\Gamma_{\rm abs}}{1- \Gamma} \right]. \label{eq: BruntVaisala}
\end{equation}

Here, it is implied that the total radiation force is provided by either electron scattering, with a constant opacity $\Gamma_{e}$, or 
an additional absorption term $\Gamma_{\rm abs}$. $H$ is the effective scale height and $a$ is the speed of sound. The additional opacity 
is described by a Kramers-like law:

\begin{align}
    \Gamma_e &= \frac{\rho}{f_{g,r}}\frac{F_r}{c} \kappa_0, \\
    \Gamma_{\rm abs} &= \frac{\rho}{f_{g,r}}\frac{F_r}{c} \kappa_1 \rho^n T^{-q},  \\
    \Gamma &= \Gamma_e + \Gamma_{\rm abs}. \label{eq: Gamma_castor}
\end{align}
%
Here, $\kappa_0$ is the Thompson opacity, which is assumed to be constant with a value of $\approx 0.2  \, \rm g^{-1} \, cm^2$ for the hydrogen-free, fully ionized gas in a WR-atmosphere. $\kappa_1$, $n$, and $q$ are the parameters describing the Kramer's law for absorption opacity. In their analysis, \cite{Castor2004} did not take into account any form of line driving. Conveniently, in our models, the radiation force close to the stellar core (where structures are first generated) is dominated by the Rosseland mean opacity and not by the line driving (Fig. \ref{fig: scat_gamma}). 

\begin{figure*}[t]
    \centering
    \includegraphics[width=\linewidth]{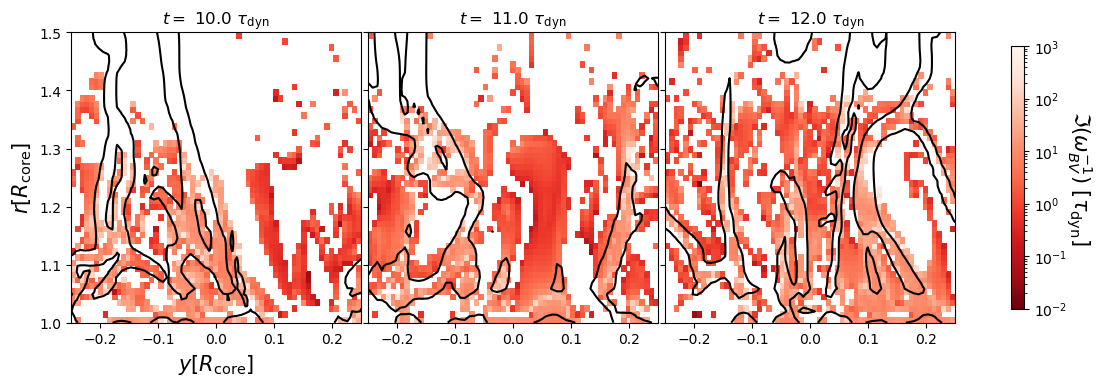}
      \caption{
      Convectively unstable zones in a series of snapshots from the 2D $\Gamma 3$ model.
   The snapshots are taken $0.5 \tau_{\rm dyn}$ apart, well after the transition of the initial conditions. The color map represents the inverse of the imaginary part of the radiation-modified Brunt-V\"ais\"al\"a frequency. Overplotted on top in black is the contour representing where the relative density is equal to one.} 
         \label{fig: BruntVaisala}
    \end{figure*}

By fitting the OPAL tables' opacity to Eq. \eqref{eq: Gamma_castor}, $n$ and $q$ can be retrieved at every cell in one of the snapshots of the 2D model. Ignoring $\kappa^{\rm line}$,  the Brunt-V\"ais\"al\"a frequency can be calculated using Eq. \eqref{eq: BruntVaisala}. $\omega_{\rm BV}$ is indeed imaginary when $\omega^2_{\rm BV}<0$ which occurs when $1/(n+q) < \Gamma_e$. In Fig. \ref{fig: BruntVaisala}, the instability growth time, given by the inverse of the imaginary part of $\omega_{\rm BV}$, is plotted for several snapshots, each a dynamical timescale apart. Here, we only focus on the lower parts of the simulation, from $r = 1 R_{\rm c}$ to  $r = 1.5 R_{\rm c}$ 
(this is the region where the analysis is mostly valid because $\kappa^{\rm line} << \kappa^{\rm OPAL}$), which is where the structures develop in our simulations, as can be seen in Fig. \eqref{fig: timeseries_rho_v}.

As can be seen from Fig. \ref{fig: BruntVaisala}, most of the lower atmosphere is convectively unstable due to the effects of radiation pressure. Moreover, for certain regions, the growth time of the instability is indeed small enough to trigger structure formation on the timescales simulated in our models. This instability is likely what  drives the formation of structure close to the WR-core in the models presented in this paper. 

\end{appendix}

\end{document}